\documentclass[5p,times]{elsarticle}

\usepackage{inconsolata}
\usepackage[T1]{fontenc}
\usepackage{booktabs,multicol,multirow,subfig,url}
\usepackage[table,xcdraw]{xcolor}

\usepackage[lined,ruled]{algorithm2e}
\SetAlFnt{\small}

\usepackage{listings}
\DeclareCaptionFormat{mylst}{\hrule\vspace{2pt}#1#2#3}
\usepackage{tikz}
\usetikzlibrary{matrix,positioning,shapes.geometric,arrows}

\usepackage{siunitx}

\begin{document}

\title{
  A SIMD algorithm for the detection of epistatic interactions of any order
}

\author[address1]{Christian Ponte-Fernández\corref{cor1}}
\ead{christian.ponte@udc.es}
\author[address1]{Jorge González-Domínguez}
\author[address1]{María J. Martín}

\cortext[cor1]{Corresponding author}
\address[address1]{Universidade da Coruña, CITIC, Computer Architecture Group,
  15071 A Coruña, Spain}

\begin{abstract}
Epistasis is a phenomenon in which a phenotype outcome is determined by the
interaction of genetic variation at two or more loci and it cannot be attributed
to the additive combination of effects corresponding to the individual loci.
Although it has been more than 100 years since William Bateson introduced this
concept, it still is a topic under active research. Locating epistatic
interactions is a computationally expensive challenge that involves analyzing an
exponentially growing number of combinations. Authors in this field have
resorted to a multitude of hardware architectures in order to speed up the
search, but little to no attention has been paid to the vector instructions that
current CPUs include in their instruction sets. This work extends an existing
third-order exhaustive algorithm to support the search of epistasis interactions
of any order and discusses multiple SIMD implementations of the different
functions that compose the search using Intel AVX Intrinsics. Results using the
GCC and the Intel compiler show that the 512-bit explicit vector implementation
proposed here performs the best out of all of the other implementations
evaluated. The proposed 512-bit vectorization accelerates the original
implementation of the algorithm by an average factor of 7 and 12, for GCC and
the Intel Compiler, respectively, in the scenarios tested.
\end{abstract}

\begin{keyword}
Epistasis, Genetic Interaction, SIMD, Vectorization, AVX\@.
\end{keyword}

\maketitle

\section{Introduction}\label{sec:introduction}

Epistasis is the interaction of genetic variation at two or more loci during the
expression of a phenotype that cannot be attributed to the additive combination
of effects corresponding to the individual loci~\cite{churchill_epistasis_2013}.
It is a phenomenon present in many plant~\cite{he_genome_2017,
jiang_quantitative_2017}, animal~\cite{banerjee_genome_2020,ruiz_evidence_2017}
and human~\cite{meijsen_using_2018,wollstein_novel_2017} traits. Because of its
importance, epistasis detection has been, and currently is, a topic under active
research.

Besides its biological implications, epistasis also represents a computational
challenge: locating a combination of features (or loci) that can correctly
classify the samples (or individuals) in different groups attending to its
phenotype, between all of the possible combinations of features. For this
reason, a multitude of methods have been proposed in order to solve this
problem, many of which refrain from exploring every combination and, instead,
implement non-exhaustive alternatives ranging from greedy algorithms to machine
learning techniques (see, for instance, \citep{meijsen_using_2018,
kim_towards_2020,shang_epiminer_2014,sun_epiaco_2017,wang_bayesian_2015}). In
a previous study~\cite{ponte-fernandez_evaluation_2020}, we compared the
performance of all epistasis detection methods published during the last decade
that offer an implementation available to the scientific community, when
locating epistatic interactions of different orders. The study concludes that,
despite the rich variety of methods, only the exhaustive approaches (those which
explore every combination of loci up to a certain combination size) can reliably
identify interactions with no marginal effects.

Given the computational complexity of finding epistatic interactions, authors in
this field have resorted to a multitude of different architectures to perform
this task, which include CPUs~\cite{wan_boost_2010,meijsen_using_2018,
campos_heterogeneous_2020}, clusters of CPUs~\cite{martinez_fast_2018,
ponte-fernandez_fast_2019}, GPUs~\cite{gonzalez_gpu_2015,nobre_exploring_2020,
campos_heterogeneous_2020}, clusters of GPUs~\cite{ponte-fernandez_fast_2019}
and other accelerators~\cite{wienbrandt_fpga_2014,luecke_fast_2015}. Most modern
CPU architectures, if not all, include Vector Processing Units (VPUs) in their
processing cores. Exploiting all the resources available in a core is key in
order to achieve the maximum performance. Although compilers incorporate
automatic vectorization techniques to exploit the VPUs in programs that do not
make explicit use of them, they show limitations on what can be automatically
vectorized, and the performance obtained is not always the optimal, as it will
be later seen. Vectorization has already been successfully employed in other
bioinformatic applications such as local sequence
alignment~\cite{galvez_blvector_2021,rucci_swimm_2019} or genome and metagenome
distance estimation~\cite{yin_rabbitmash_2020}. However, despite the potential
that an efficient use of these units offers, none of the works previously
mentioned, with the exception of \citep{campos_heterogeneous_2020}, consider
their use to further accelerate the epistasis search. In
\citep{campos_heterogeneous_2020} Campos \textit{et al.} include a couple of
Intel Advanced Vector Extensions (AVX) Intrinsics to parallelize the bitwise
\textit{AND} and \textit{AND-NOT} operations, used during the computation of the
genotype frequencies corresponding to a particular loci combination.

In this work, we detailedly explore the intricacies of the SIMD parallelization
of an exhaustive epistasis detection algorithm to find interactions of any
order, looking to maximize the performance per core in modern CPUs. To do this,
we manually implement an epistasis search explicitly using vector instructions,
and compare it with the performance achieved by the GCC and Intel compilers
automatically vectorizing the same operations. Starting from an existing method
for third-order epistasis detection (MPI3SNP~\cite{ponte-fernandez_fast_2019}),
Section~\ref{sec:exh_search} describes how this method can be expanded to
support loci combinations of any size. We selected MPI3SNP due to its good
performance shown when locating epistasis~\cite{ponte-fernandez_evaluation_2020}
and the simplicity of the algorithm implemented. Section~\ref{sec:vectorization}
explains how each of the operations can be vectorized through AVX Intrinsics
using the \textit{AVX2} and \textit{AVX512BW} extensions.
Section~\ref{sec:evaluation} presents the experimental evaluation of the
vectorized algorithm, comparing the performance achieved by the explicit vector
implementations using AVX Intrinsics with the performance achieved by the
automatic vectorization that compilers offer and the original performance of the
MPI3SNP implementation. Finally, Section~\ref{sec:conclusions} discusses the
conclusions extracted from this study, reflects on its limitations and comments
on some lines of future work.

\section{Exhaustive Epistasis Detection}\label{sec:exh_search}

This work explores the extension of the exhaustive third-order epistasis
detection algorithm used by MPI3SNP~\cite{ponte-fernandez_fast_2019} to support
the detection of epistasis interactions of any order, as well as the
optimization of its implementation from the perspective of its single-thread
performance.

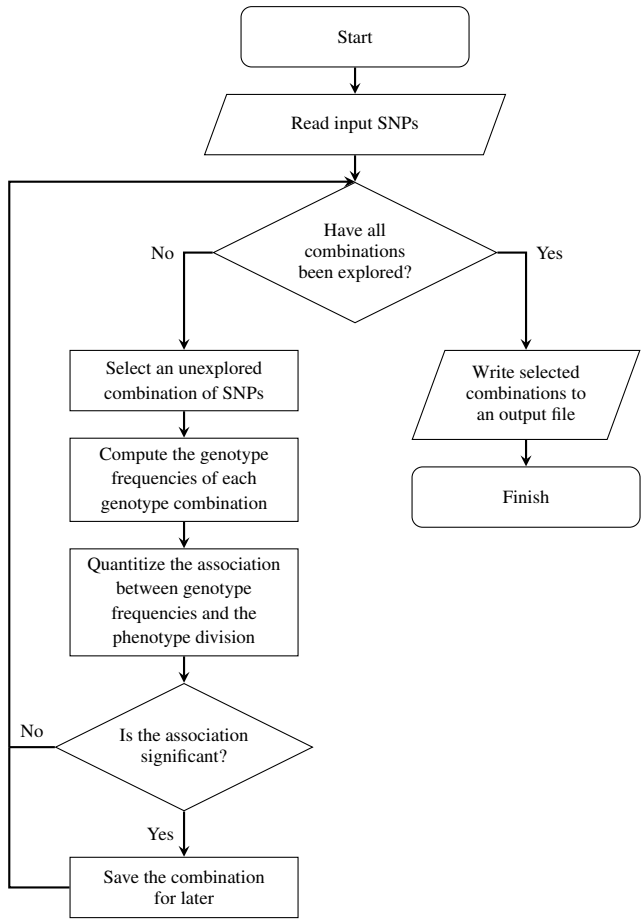
\begin{figure}
    \centering
    \begin{tikzpicture}[
        node distance=0.35cm,
        font=\scriptsize,
        startstop/.style={
            rectangle,
            rounded corners,
            minimum width=3cm,
            minimum height=0.8cm,
            text centered,
            draw=black,
        },
        io/.style={
            trapezium,
            trapezium left angle=70,
            trapezium right angle=110,
            minimum width=3cm,
            minimum height=0.8cm,
            text centered,
            draw=black,
        },
        process/.style={
            rectangle,
            minimum width=3cm,
            minimum height=0.8cm,
            text centered,
            draw=black,
        },
        decision/.style={
            diamond,
            aspect=2,
            text centered,
            draw=black,
        },
        arrow/.style={
            thick,
            ->,
            >=stealth
        }
    ]
        \node (start) [startstop]
            {Start};
        \node (in1) [io, below=of start.south]
            {Read input SNPs};
        \node (dec1) [decision, below=of in1.south]
            {\shortstack{Have all\\combinations\\been explored?}};
        \node (pro1a) [process, below=of dec1.south, xshift=-2.25cm]
            {\shortstack{Select an unexplored\\combination of SNPs}};
        \node (pro1b) [process, below=of pro1a.south]
            {\shortstack{Compute the genotype\\frequencies of each\\genotype
            combination}};
        \node (pro1c) [process, below=of pro1b.south]
            {\shortstack{Quantitize the association\\between genotype\\
            frequencies and the\\phenotype division}};
        \node (dec2) [decision, below=of pro1c.south]
            {\shortstack{Is the association\\significant?}};
        \node (pro3) [process, below=of dec2.south, yshift=-0.25cm]
            {\shortstack{Save the combination\\for later}};
        \node (out1) [io, below=of dec1.south, xshift=2.25cm]
            {\shortstack{Write selected\\combinations to\\an output file}};
        \node (finish) [startstop, below=of out1.south]
            {Finish};
        \draw [arrow] (start) -- (in1);
        \draw [arrow] (in1) -- (dec1);
        \draw [arrow] (dec1) -| node[anchor=east] {No} (pro1a.north);
        \draw [arrow] (dec1) -| node[anchor=west] {Yes} (out1.north);
        \draw [arrow] (pro1a) -- (pro1b);
        \draw [arrow] (pro1b) -- (pro1c);
        \draw [arrow] (pro1c) -- (dec2);
        \draw [arrow] (dec2) -- node[anchor=east] {Yes} (pro3);
        \coordinate [left=of dec2.west, xshift=-0.25cm] (aux1);
        \draw [thick] (dec2.west) -- node[anchor=south] {No} (aux1);
        \draw [thick] (pro3.west) -| (aux1);
        \draw [arrow] (aux1) |- (dec1.north);
        \draw [arrow] (out1.south) -- (finish);
    \end{tikzpicture}
    \caption{
        Flowchart of a typical exhaustive epistasis detection method.
    }\label{fig:exhaustive_search_flowchart}
\end{figure}

Exhaustive epistasis detection methods, despite their differences in
implementation, follow a common set of steps summarized in the flowchart shown
in Figure~\ref{fig:exhaustive_search_flowchart}. The key elements of the program
flow are:

\begin{enumerate}
  \item The enumeration of all combinations of loci, without repetition, for a
        set of loci and up to a certain combination size.
  \item The computation of the different allele combination frequencies for
        each combination of loci.
  \item The quantification of the association between the division of the
        individuals in groups (such as cases and controls), and the differences
        in genotype frequencies between those groups, for each combination.
\end{enumerate}

In this work, a depth-first algorithm is used for the enumeration of
combinations of any given size, genotype and contingency tables are used for the
computation of the different allele frequencies and the Mutual information (MI)
method is employed for the quantification of the association between groups.

Genotype tables were first introduced by Wan \textit{et al.} in BOOST~\cite{
wan_boost_2010}, and since then have been adopted by a variety of epistasis
detection works, including MPI3SNP~\cite{ponte-fernandez_fast_2019} (some other
examples are \citep{campos_heterogeneous_2020,nobre_exploring_2020,
guo_cloud_2014}). Genotype tables are a data structure used to represent the
genotype information of a particular locus or combination of loci, following a
binary format that allows for the combination with further loci using binary
operators exclusively. This work uses Single Nucleotide Polymorphisms (SNPs) as
the input genotype information. An SNP represents a specific locus in the genome
where at least 1\% of the population presents a genomic variation.

Genotype tables, once calculated, can be transformed into contingency tables
that represent the allele combination frequencies of all individuals by simply
counting bits on the genotype table rows. Wan \textit{et al.} present the
genotype and contingency tables as a single operation. However, when exploring
combinations of more than two loci, as is the case here, it is convenient to
separate both operations since many combinations share a common number of loci
with one another, and thus many intermediate genotype tables can be used to
compute the contingency tables of different combinations.
Subsections~\ref{ss:method_gt}~and~\ref{ss:method_ct} cover the calculation of
genotype and contingency tables for combinations of variable size, respectively.

Once the contingency table of a particular loci combination is calculated, its
association with the phenotype of interest can be measured using MI\@. It is a
metric from Information Theory that measures the amount of information
obtainable from one variable through the observation of another.
Subsection~\ref{ss:method_mi} covers the calculation of this metric. Lastly,
after all the operations are defined, Subsection~\ref{ss:nonseg_exploration}
describes the algorithm that combines these operations to exhaustively explore
every combination of loci for a particular combination size and locate the ones
most associated with the phenotype.

\subsection{Genotype table calculation}\label{ss:method_gt}

Genotype tables are data structures used to encode the genotype information
following a binary format. Genotype tables group individuals by their phenotype
class into two separate subtables, one for cases and another for controls. These
subtables contain as many rows as genotypes each individual can have, and as
many bits in each row as individuals represented in the table. When representing
a single SNP of a biallelic population, as is the case with human populations,
each individual can have three different genotypes, encoded in three different
rows: homozygous dominant, heterozygous and homozygous recessive. A 0 or a 1 in
a particular position of a row indicates the absence or presence of the genotype
corresponding to that row for the individual represented in that position,
respectively. Using this representation, the information of a singular SNP for
\(m\) individuals can be encoded in \(3m\) bits. Fig.~\ref{fig:gt_example}
includes two examples of genotype tables for two different SNPs \(i\) and \(j\),
both of which include 32 individuals and are encoded using 96 bits.

\begin{figure}
  \centering

  \tikzset{
    title/.style={
      font=\bfseries\small
    },
    table/.style={
      matrix of nodes,
      row sep=-\pgflinewidth,
      column sep=-\pgflinewidth,
      nodes={
        rectangle,
        draw=black,
        align=center,
        font=\small\ttfamily
      },
      text depth=0em,
      nodes in empty cells,
      column 1/.style={
        nodes={
          text width=2em,
          font=\scriptsize,
          draw=none,
          fill=none,
          align=right}
      },
      column 2/.style={
        every even row/.style={
          nodes={
            fill=gray!20
          }
        }
      },
      column 3/.style={
        every even row/.style={
          nodes={
            fill=gray!20
          }
        }
      },
      row 1/.style={
        nodes={
          fill=none,
          draw=none,
          font=\small
        }
      },
      row 6/.style={
        nodes={
          text height=1em
        }
      }
    }
  }
  \begin{tikzpicture}
    \matrix[table,text width=8em] (a)
    {
                & Cases             & Controls          \\
      \(i_{1}\) & 01011010 00001111 & 01010010 10001111 \\
      \(i_{2}\) & 00000101 00100000 & 10000100 01100000 \\
      \(i_{3}\) & 10100000 11010000 & 00101001 00010000 \\
    };
    \node[title] (a_title) [above left of=a-1-1,right,node distance=2em]
      {Genotype table of the SNP \textit{i}:};

    \matrix[table,text width=8em,below=2em of a] (b)
    {
                & Cases             & Controls          \\
      \(j_{1}\) & 00101011 10010101 & 00100001 00111001 \\
      \(j_{2}\) & 10010000 00000000 & 11001010 10000100 \\
      \(j_{3}\) & 01000100 01101010 & 00010100 01000010 \\
    };
    \node[title] (b_title) [above left of=b-1-1,right,node distance=2em]
      {Genotype table of the SNP \textit{j}:};

    \matrix[table,text width=8em,below=2em of b] (c)
    {
                           & Cases             & Controls          \\
      \(i_{1} \cap j_{1}\) & 00001010 00000101 & 00000000 00001001 \\
      \(i_{1} \cap j_{2}\) & 00010000 00000000 & 01000010 10000100 \\
      \(i_{1} \cap j_{3}\) & 01000000 00001010 & 00010000 00000010 \\
      \(i_{2} \cap j_{1}\) & 00000001 00000000 & 10000000 00000000 \\
                           & \vdots            & \vdots            \\
      \(i_{3} \cap j_{3}\) & 00000000 01000000 & 00000000 00000000 \\
    };
    \node[title] (c_title) [above left of=c-1-1,right,node distance=2em]
      {Genotype table of SNP \textit{i} {\texttimes} SNP \textit{j}:};
  \end{tikzpicture}

  \caption{
    Example of two genotype tables of two different SNPs, \(i\) and \(j\), for
    sixteen cases and controls, and the combined genotype table of the two SNPs.
  }\label{fig:gt_example}
\end{figure}

Genotype tables simplify the operation of combining different SNPs, as they can
also be used to represent the combination of multiple genotypes. Considering
that each row of the table represents the presence of a particular genotype in
every individual as set bits, one can find which individuals have a particular
genotype combination by calculating the intersection between the corresponding
table rows. Therefore, obtaining the genotype distribution of a particular SNP
combination involves combining the rows of the different tables and calculating
their intersection via bitwise \textit{AND} operations, for the case and control
subtables separately. This procedure can be used to combine as many SNPs as
necessary, resulting in a genotype table containing \(3^s\) rows and \(3^s m\)
bits in total, with \textit{s} being the number of SNPs in combination and
\textit{m} the number of individuals represented. Fig.~\ref{fig:gt_example} also
includes an example of this, in which the two previous example SNPs \(i\) and
\(j\) are combined into a genotype table containing a total of 9 rows.

Note that in an exhaustive exploration of combinations up to a certain size,
smaller combinations are bound to appear in many larger ones. In this work, to
maximize the reuse of tables and, thus, the performance, instead of combining
numerous SNPs at once, genotype tables are constructed following an iterative
process, adding one SNP to an existing table in each step. Additionally, and in
contrast with the original method in BOOST~\cite{wan_boost_2010}, we are
decoupling the representation of the genotype information in tables from
contingency tables representing numeric genotype frequencies, as the latter
loses the information of singular individuals and the frequencies are not needed
in every step of the algorithm.

Listing~\ref{lst:gt_base} shows a candidate C++ implementation of this operation
in function \texttt{combine}, taking as arguments a genotype table representing
the combination of any number of SNPs, a genotype table of a singular SNP and a
genotype table where the results will be stored. Note that the template argument
\texttt{\detokenize{uint64_t}}, common to all genotype table classes, indicates
the type used to store the binary information
(Lines~\ref{code:gt_1}--\ref{code:gt_2}). Since the \detokenize{x86_64}
instruction set operates with 64-bit integers, this type is ideal to hold the
genotype information, each value representing the information of 64 individuals,
and each computation operating with 64 individuals at once. The function calls
the \texttt{combine\_subtable} subroutine twice to combine each of the two
subtables for cases and controls. This function consists of three nested
\textit{for} loops, the two outermost loops
(Lines~\ref{code:gt_3}~and~\ref{code:gt_4}) combine the different rows of the
input genotype tables. The innermost loop (Line~\ref{code:gt_5}) iterates over
the different \texttt{\detokenize{uint64_t}} values of the selected rows,
reading one value from each input table, calculating its intersection and
storing the result in the output table. The computational time complexity of
this operation is \(O(3^s m)\), where \(m\) is the number of individuals in the
data and \(s\) is the size of the combination.

\subsection{Contingency table calculation}\label{ss:method_ct}

Contingency tables represent the frequency distribution of two discrete
variables from a number of observations. For this domain of application, the
variables represented in the contingency tables are the phenotype and genotype
variation. Constructing these tables from the genotype table representation is
direct, as individuals are already segregated into different subtables according
to the case and control groups, and different rows according to their genotype
information. Obtaining the frequency distribution can be done by counting how
many bits are set to one for each row. This operation is known as population
count (or, in short, \textit{popcount}), and most of the current processors
provide hardware support for this operation.

Separating the genotype table and contingency table calculations into completely
different operations may seem convenient at first glance, as this allows for
genotype tables common to multiple SNP combinations to be reused. However, if we
compute them entirely separate, the resulting values from the computation of the
genotype table are stored in memory just to be brought back immediately after,
in order to compute the corresponding contingency table. Because of this, it is
more convenient to compute the rows of the last-level genotype table and,
instead of storing them as usual, perform the \textit{popcount} operation in a
single step, saving us from several load and store instructions.

Listing~\ref{lst:ct_base} shows a C++ implementation for this operation, very
similar to the genotype table calculation (Listing~\ref{lst:gt_base}), showing
that the function \texttt{combine\_and\_popcnt} consists of two calls to the
subroutine \texttt{popcnt\_subtable} to compute the contingency subtables of the
two new genotype subtables, and this subroutine consists of the same three
nested loops. However, instead of immediately storing the multiple
\texttt{uint64\_t} values resulting from the bitwise \textit{AND} operations,
the \textit{popcount} operation is called, the results of the same row are
summed up  in a single \texttt{uint32\_t} value and the sum is stored in the
contingency table (Lines~\ref{code:ct_2}--\ref{code:ct_3}). In contrast with the
genotype table class, the contingency table class uses the
\texttt{\detokenize{uint32_t}} type (passed as a template argument in
Line~\ref{code:ct_1}) to represent the total count of individuals having a
particular genotype because it can contain a large enough integer, and for the
convenience of matching the size of a \texttt{float} value which will be useful
later during the vectorization of the MI operation. The computational time
complexity of this operation is also \(O(3^s m)\), with \(m\) being the number
of individuals in the data and \(s\) the size of the combination.

\subsection{Mutual Information calculation}\label{ss:method_mi}

MI quantifies the degree of association between the genotype distribution of
cases and controls and the phenotype distribution obtained from the previously
calculated contingency table. MI has shown a very good detection power in our
previous comprehensive study~\cite{ponte-fernandez_evaluation_2020}, and the
presence of low frequencies in the data, which become more prevalent as we move
towards larger combination sizes, do not seem to be a problem.

MI is a metric from Information Theory that measures the amount of information
obtainable from one variable through the observation of another one. Taking the
genotype and phenotype variability as two random variables \textit{X} and
\textit{Y}, MI can be obtained as:

\begin{equation}
  MI(X; Y) = H(X) + H(Y) - H(X,Y)
\end{equation}

where \(H(X)\) and \(H(Y)\) are the marginal entropies of the two variables, and
\(H(X,Y)\) is the joint entropy. Marginal entropies of a single and two
variables are defined as:

\begin{equation}
  H(X) = - \sum_{x \in X} p(x)\log p(x)
\end{equation}

\begin{equation}
  H(X,Y) = - \sum_{x, y} p(x, y)\log p(x, y)
\end{equation}

with \(p(x)\) representing the probability of the random variable \(X\) taking
the value \(x\), \(p(y)\) the probability of the random variable \(Y\) taking
the value \(y\), and \(p(x,y)\) the joint probability of both events. These
probabilities can be obtained directly from the contingency table as the
division between the number of occurrences and the number of total observations.

Listing~\ref{lst:mi_base} shows a base C++ implementation of the function to
calculate the MI from an existing contingency table. It computes \(H(X,Y)\) and
\(H(X)\) in a single \textit{for} loop (Lines~\ref{code:mi_3}--\ref{code:mi_4}
and~\ref{code:mi_5}--\ref{code:mi_6}, respectively). The loop includes three
\textit{if} branches to avoid computing the logarithm of 0, which would lead to
an undefined product of \(0 \times -{\infty}\), resulting in a \texttt{NaN}
value. \(H(Y)\) and the inverse of the number of individuals (\texttt{iinds})
are provided as function arguments because they are independent of the genotype
distribution of individuals, and thus can be calculated just once outside the
function (Lines~\ref{code:mi_1}~and~\ref{code:mi_2}). The MI function operates
with \texttt{float} types since single-precision floating point numbers offer
enough numerical precision to represent the MI values. The time complexity of
this operation is \(O(3^s)\), where \(s\) is the size of the combination
represented in the input contingency table.

\subsection{Combinatory exploration algorithm}\label{ss:nonseg_exploration}

\begin{algorithm}[t]
  \DontPrintSemicolon{}
  \KwIn{\;
    \(d\): Data set containing \(n\) genotype tables, each representing a
    single SNP for \(m\) individuals\;
    \(o\): Order of the interactions to identify
  }
  \KwOut{\;
    \(v\): Vector of combinations of \(o\) SNPs, and their associated MI
    value
  }
  {\BlankLine}
  {\nl} Allocate an array \(g\) of \(o-2\) genotype tables, for sizes between 1
    and \(o-1\) \;
  {\nl} Allocate a contingency table \(c\) of size \(o\) \;
  {\nl} Create an empty stack \(s\) \;
  {\nl} Calculate the marginal entropy \(H(Y)\) \;\label{code:alg_1}
  {\nl} \(inv\_inds \leftarrow 1 / m\) \;\label{code:alg_2}
  {\nl} \For{\(i \leftarrow 0\) {\KwTo} \(n\)}{
    {\nl} \(g[1] \leftarrow d[i]\) \;\label{code:alg_3}
    {\nl} \For{\(j \leftarrow i + 1\) {\KwTo} \(n\)}{
        {\nl} Push the pair \( \{i,j\} \) into the stack \(s\)
          \;}\label{code:alg_4}
    {\nl} \While{\(s\) is not empty}{
      {\nl} Pop the combination \( \{c_1,\ldots,c_k\} \) from the top of the
        stack \(s\) \;\label{code:alg_5}
      {\nl} \If{\(k < o\)}{
        {\nl} \(g[k] \leftarrow combine(g[k - 1], d[c_k])\) \;\label{code:alg_6}
        {\nl} \For{\(j \leftarrow c_k + 1\) {\KwTo} \(n\)}{\label{code:alg_7}
          {\nl} Push the new combination \( \{c_1,\ldots,c_k,j\} \) into the
            stack \(s\) \;
        }\label{code:alg_8}
      }
      {\nl} \Else{
        {\nl} \(c \leftarrow combine\_and\_popcnt(\)
        \(g[o-1], d[c_k])\) \;\label{code:alg_9}
        {\nl} \(mi\_val \leftarrow MI(c, H(Y), inv\_inds)\) \;
        {\nl} Add \( \{c_1,\ldots,c_k\} \) and \textit{mi\_val} to the vector of
          results \textit{v}\;\label{code:alg_add_res}
      }
    }
  }
  \caption{Non-segmented exploration}\label{alg:nonseg}
\end{algorithm}

Algorithm~\ref{alg:nonseg} shows the pseudocode of a depth-first exploration
algorithm which relies on the genotype table, contingency table and MI
functions previously defined to combine the different SNPs of the input data
and assess the degree of association between the SNP combinations and the
phenotype of study. The key element of this algorithm is that it iterates over
all the combinations in a depth-first manner with the help of a stack. This is
fundamental to prevent multiple calculations of the same genotype table, since
combinations may share a common set of SNPs with other combinations. When the
combination space is explored depth-first, we exhaust all combinations starting
with a particular prefix (and its corresponding genotype table) before moving
onto the next one.

The arguments to this routine are the data set \(d\) containing the genotype
tables of all individual SNPs in the data, the order \(o\) of the interactions
to locate and a vector \(v\) where the results will be returned. In the first
three lines, the function starts by allocating enough space for an array \(g\)
of \(o-2\) genotype tables of size \(1\) to \(o-1\), a contingency table \(c\)
for combinations of the target size \(o\) and a stack \(s\) of combinations of
SNP indexes. Before starting the exploration, the function computes the inverse
of the number of individuals \(inv\_inds\), and the entropy of the phenotype
variability \(H(Y)\) (Lines~\ref{code:alg_1}~and~\ref{code:alg_2}), the two
arguments of the MI function common to all combinations.

After that, the function starts to loop through all SNPs, exploring all of the
combinations starting with that SNP before moving onto the next one. To do this,
the genotype table of the SNP \(i\) is copied in \(g[1]\), and all combinations
of two SNPs starting with that one are pushed into the stack
(Lines~\ref{code:alg_3}--\ref{code:alg_4}). Then, using a \textit{while} loop,
the combinations of the stack are processed until it is emptied. In each
iteration, the top combination \( \{c_1,\ldots,c_k\} \) of the stack is popped
(Line~\ref{code:alg_5}). If \(k\) is smaller than the target interaction order
\(o\), its corresponding genotype table is computed from the genotype table of
its prefix (stored in the array \(G\)) and the table of the last SNP \(c_k\)
(Line~\ref{code:alg_6}). Then, all subsequent combinations starting with \(
\{c_1,\ldots,c_k\} \) are pushed into the stack
(Lines~\ref{code:alg_7}--\ref{code:alg_8}). Otherwise, if \(k\) is equal to
\(o\), its contingency table and MI are computed, and the result is stored in
the vector of results \(v\) (Lines~\ref{code:alg_9}--\ref{code:alg_add_res}).

For simplicity, in the pseudocode, we are appending all combinations with its MI
to the vector of results, although in the actual implementation only the
combinations with the highest MI value are retained in the vector. The
computational time complexity of this algorithm is \(O\left({{n}\choose{o}} 3^o
m\right)\), where \(n\) is the number of SNPs in the data, \(m\) is the number
of individuals in the data, and \(o\) is the size of the combinations explored.

\section{SIMD Implementation}\label{sec:vectorization}

This section covers the explicit vectorization of the operations presented in
the previous section, using 256-bit and 512-bit AVX Intrinsics from the
\textit{AVX2} and \textit{AVX512BW} extensions. This section also addresses
several optimizations introduced both to the individual operations and the
general exhaustive algorithm to improve the performance of the vectorized
codes.

The \textit{AVX2} vector extension was first introduced with the Intel Haswell
microarchitecture (2013) while the \textit{AVX512BW} extension first appeared in
the Skylake-X processors (2017) of the Skylake microarchitecture. In this work,
these two vector extensions are used not only to optimize the runtime of the
epistasis detection tool on a long list of CPUs, but also to compare the
performance that the two vector widths offer.

\subsection{Vectorization of the genotype table calculation}

The function \texttt{combine\_subtable} shown in Listing~\ref{lst:gt_base} is
the one implementing the computation of a genotype subtable from two previous
subtables, and thus our target for vectorization. In this function, we can
identify a vectorization opportunity at the innermost loop, where the
intersection of two rows from two tables is calculated by performing as many
bitwise \textit{AND} operations as values contained in the row
(Line~\ref{code:gt_6}). This operation is already exploiting the
data-parallelism that 64-bit operations offer, as the information of a singular
individual is stored in a single bit of the data type
\texttt{\detokenize{uint64_t}}. With the introduction of 256 and 512-bit AVX
instructions, the throughput of this operation can be multiplied.

Listing~\ref{lst:gt_avx2} shows the implementation using 256-bit AVX Intrinsics
from \textit{AVX2}. For simplicity, we assume that the number of bytes in a row
of the genotype table is divisible by the vector unit width. This is achieved by
padding the rows with zeros if the number of individuals is not divisible by the
width of the vector unit, and it will not influence the result of the following
\textit{popcount} operation. The new figure replaces the C++ code corresponding
to the two array accesses, the \textit{AND} and the store operations with AVX
loads, \textit{AND}s and store intrinsics. With just the introduction of the AVX
Intrinsics, there is a front-end bound problem in which the CPU wastes many
clock cycles waiting for instructions to be fetched. Therefore, to correct this
behaviour, the middle loop was unrolled completely so that the three rows from
the second genotype table are processed concurrently.

The 512-bit vector implementation using intrinsics from the \textit{AVX512BW}
extension is almost identical to the one shown in Listing~\ref{lst:gt_avx2}, and
thus it was omitted. The only differences are the name of the functions that
implement the same operations for a 512-bit width, the types that these
operations use and the step of the innermost loop, which doubles the one used in
the 256-bit implementation.

\subsection{Vectorization of the contingency table calculation}

\begin{table*}[t]
  \caption{
    Elapsed time, in seconds, during the computation of a single contingency
    table using different operation widths and \textit{popcount}
    implementations. The table highlights with green background the best times
    using the \textit{AVX512BW} extension, and with red text the best times
    using the \textit{AVX2} extension.
  }\label{tbl:mula_popcnt_results}
  \centering
  \begin{tabular}{lllllllll}
    \toprule
    \multirow{2}{*}[-1pt]{\bfseries\shortstack[l]{\texttt{AND}\\width}}
      & \multirow{2}{*}[-1pt]{\bfseries\shortstack[l]{\texttt{POPCNT}\\width}}
      & \multirow{2}{*}[-1pt]{\bfseries\shortstack[l]{\texttt{POPCNT}\\algorithm}}
      & \multicolumn{6}{c}{\bfseries Individuals count} \\
      &
      &
      & \textbf{256}
      & \textbf{512}
      & \textbf{1024}
      & \textbf{2048}
      & \textbf{4096}
      & \textbf{8192} \\
    \midrule
    512
      & 512
      & harley seal
      & \num{8.60e-07}
      & \num{8.60e-07}
      & \num{1.02e-06}
      & \num{1.31e-06}
      & \num{1.88e-06}
      & \cellcolor[HTML]{C6EFCE}{\num{1.60e-06}} \\
    512
      & 512
      & lookup
      & \num{2.47e-07}
      & \cellcolor[HTML]{C6EFCE}{\num{2.47e-07}}
      & \cellcolor[HTML]{C6EFCE}{\num{3.39e-07}}
      & \cellcolor[HTML]{C6EFCE}{\num{5.57e-07}}
      & \cellcolor[HTML]{C6EFCE}{\num{9.63e-07}}
      & \num{1.78e-06} \\
    512
      & 256
      & cpu
      & \num{4.04e-07}
      & \num{4.04e-07}
      & \num{6.91e-07}
      & \num{1.18e-06}
      & \num{2.20e-06}
      & \num{4.19e-06} \\
    512
      & 256
      & harley seal
      & \num{7.59e-07}
      & \num{7.59e-07}
      & \num{1.05e-06}
      & \num{1.61e-06}
      & \num{1.72e-06}
      & \num{2.82e-06} \\
    512
      & 256
      & lookup
      & \num{2.98e-07}
      & \num{2.98e-07}
      & \num{4.64e-07}
      & \num{7.66e-07}
      & \num{1.40e-06}
      & \num{2.67e-06} \\
    512
      & 256
      & lookup orig.
      & \num{2.99e-07}
      & \num{2.99e-07}
      & \num{4.57e-07}
      & \num{8.01e-07}
      & \num{1.46e-06}
      & \num{2.81e-06} \\
    512
      & 64
      & popcnt movdq
      & \num{3.02e-07}
      & \num{3.02e-07}
      & \num{5.19e-07}
      & \num{9.99e-07}
      & \num{1.86e-06}
      & \num{3.60e-06} \\
    512
      & 64
      & popcnt un.\ err.
      & \num{4.36e-07}
      & \num{4.36e-07}
      & \num{7.09e-07}
      & \num{1.17e-06}
      & \num{2.06e-06}
      & \num{3.82e-06} \\
    256
      & 256
      & cpu
      & \num{1.95e-07}
      & \num{2.90e-07}
      & \num{5.32e-07}
      & \num{9.38e-07}
      & \num{1.81e-06}
      & \num{3.46e-06} \\
    256
      & 256
      & harley seal
      & \num{5.08e-07}
      & \num{6.06e-07}
      & \num{7.85e-07}
      & \num{1.16e-06}
      & \num{1.14e-06}
      & \num{1.85e-06} \\
    256
      & 256
      & lookup
      & \num{2.24e-07}
      & \num{3.13e-07}
      & \num{4.65e-07}
      & {\color[HTML]{9C0006} \num{5.71e-07}}
      & {\color[HTML]{9C0006} \num{9.98e-07}}
      & {\color[HTML]{9C0006} \num{1.83e-06}} \\
    256
      & 256
      & lookup orig.
      & \num{2.15e-07}
      & \num{2.90e-07}
      & {\color[HTML]{9C0006} \num{4.56e-07}}
      & \num{7.82e-07}
      & \num{1.44e-06}
      & \num{2.75e-06} \\
    256
      & 64
      & popcnt movdq
      & \cellcolor[HTML]{C6EFCE}{\color[HTML]{9C0006} \num{1.60e-07}}
      & {\color[HTML]{9C0006} \num{2.58e-07}}
      & \num{4.73e-07}
      & \num{8.82e-07}
      & \num{1.73e-06}
      & \num{3.36e-06} \\
    256
      & 64
      & popcnt un.\ err.
      & \num{1.86e-07}
      & \num{3.08e-07}
      & \num{5.42e-07}
      & \num{1.04e-06}
      & \num{2.02e-06}
      & \num{3.93e-06} \\
    \bottomrule
  \end{tabular}
\end{table*}

The main difference between the codes for calculating genotype and contingency
tables (Listings~\ref{lst:gt_base}~and~\ref{lst:ct_base}, respectively)
is the presence of the \textit{popcount} operation. Up until very recently, with
the introduction of the Intel Ice Lake processors, there was no AVX vector
instruction implementing a \textit{popcount}. Muła \textit{et al.},
in~\cite{mula_faster_2018}, have already explored this problem and they proposed
multiple algorithms for implementing population counts using the \textit{AVX2}
extension. Furthermore, in their Github repository~\cite{mula_github_sse}, they
have developed updated versions of the algorithms to make use of the more recent
\textit{AVX512BW} and \textit{AVX512VBMI} extensions.

Deciding which algorithm runs the fastest is not trivial and cannot be measured
in isolation, as interleaving additional loads and bitwise \texttt{AND}
operations in between \textit{popcounts} will undoubtedly affect the performance
of the function as a whole. For this reason, we implemented multiple versions of
the \texttt{combine\_{\allowbreak}and\_popcount} function and compared the
performance of each choice. Table~\ref{tbl:mula_popcnt_results} includes the
elapsed time during the computation of a contingency table for the different
implementations of the function on an Intel Xeon Gold 6240 CPU, the processor
used for the experimental evaluation in Section~\ref{sec:evaluation}, and
compiled with GCC version 8.3.0. The table considers:

\begin{enumerate}
  \item Two different vector widths for the bitwise \texttt{AND} operations: 256
        and 512 bits.
  \item Three different vector widths for the \textit{popcount} operations: 64
        bits, using the hardware \textit{popcount} instruction from the Bit
        Manipulation Instructions (BMI) extension, and 256 and 512 bits, using
        the software implementations proposed
        in~\cite{mula_faster_2018,mula_github_sse}.
  \item Six different table row widths: 256, 512, 1024, 2048, 4096 and 8192
        individuals in each row (or 32, 64, 128, 256, 512 and 1024 bytes per
        row, respectively), equal for cases and controls.
\end{enumerate}

From these results we can conclude that the fastest implementation is dependant
on the width of the genotype tables. For less than 512 individuals per subtable,
the best times are obtained by the implementations that make use of the 64-bit
hardware \textit{popcount} instruction. However, if we have more than 512
individuals, the lookup implementations for both the \textit{AVX2} and
\textit{AVX512BW} extensions offer the fastest alternative for most of the
widths tested. Taking a look at all of the epistasis studies referenced
throughout this paper, we can find that most of them consider a number of
individuals between 512 and 4096. Therefore, for this work we will use the
\textit{AVX2} and \textit{AVX512BW} implementations of the \textit{popcount}
lookup algorithm.

Vectorizing the \texttt{combine\_and\_popcount} function from
Listing~\ref{lst:ct_base} requires vectorizing its auxiliary subroutine
\texttt{popcnt\_{\allowbreak}subtable}. Starting with the \textit{AVX2}
implementation, Listing~\ref{lst:ct_avx2} shows an implementation of the
vectorized function, combining the computation of the new genotype table with
the \textit{popcount} lookup algorithm. This function includes the following
modifications to the original lookup algorithm:

\begin{enumerate}
  \item Instead of iterating over an input array as in the original popcount
        function (Lines 32--43 from file
        \texttt{popcnt-avx2-{\allowbreak}lookup.cpp}
        from~\cite{mula_github_sse}), \texttt{popcnt\_subtable} consists of
        three nested loops: the two outer ones
        (Lines~\ref{code:vct256_1}~and~\ref{code:vct256_2}) combining the
        different rows of the input genotype tables, and the two innermost loops
        (Lines~\ref{code:vct256_3}~and~\ref{code:vct256_4}) applying the
        \textit{popcount} iteration to each 256-bit word of the two selected
        rows. The first of the two innermost loops
        (Lines~\ref{code:vct256_3}--\ref{code:vct256_10}) maintains the original
        unrolling of eigth 256-bit words.
  \item Each iteration step (inlined function \texttt{iter}) reads a 256-bit
        word from each table row
        (Lines~\ref{code:vct256_5}~and~\ref{code:vct256_6}), computes the
        bitwise \texttt{AND} of the two words (Line~\ref{code:vct256_7}) and
        continues with the Muła \textit{popcount} algorithm
        (Lines~\ref{code:vct256_8}--\ref{code:vct256_9}, which correspond to
        Figure 10 from~\cite{mula_faster_2018}).
\end{enumerate}

Listing~\ref{lst:ct_avx512bw} shows the implementation of the same
\texttt{popcnt\_{\allowbreak}subtable} subroutine but using Intrinsics from the
\textit{AVX512BW} extension. The original \textit{popcount} lookup algorithm for
\textit{AVX512BW} (file \texttt{popcnt-avx512bw-lookup.cpp}
from~\cite{mula_github_sse}) is very similar to its \textit{AVX2}
implementation, with the obvious difference of not applying unrolling to its
innermost loop (Lines 39--49). Therefore, the same considerations for the
\textit{AVX2} implementation of the function apply to the \textit{AVX512BW}
algorithm: the function combines the input genotype tables using three nested
loops (Lines~\ref{code:vct512_1},\ref{code:vct512_2}~and~\ref{code:vct512_3}),
and each \textit{popcount} iteration of the Muła algorithm is preceded by two
loads that read a 512-bit word from each input genotype table
(Lines~\ref{code:vct512_4}~and~\ref{code:vct512_5}) and a bitwise AND operation
(Line~\ref{code:vct512_6}).

\subsection{Vectorization of the Mutual Information calculation}

In contrast to the two previous functions, calculating the MI of a contingency
table requires floating-point arithmetic, including multiplications, fused
multiply-adds (FMAs) and logarithms. Multiplications and FMAs are supported
natively, both for 256-bit and 512-bit vector operations, but there is no
hardware instruction that implements a logarithm. However, Intel does provide an
AVX logarithm routine through their Short Vector Math Library (SVML), an
extension to the Intel Intrinsics available only with the Intel Compiler. GCC
provides a vector implementation of the logarithm through GNU \textit{libc}'s
vector math library, available since version 2.22, although the number of vector
functions available with GNU \textit{libc} is much more limited compared to
Intel's SVML\@.

Listing~\ref{lst:mi_avx2} shows a C++ function implementing the MI computation
using AVX Intrinsics from the \textit{AVX2} extension. This code assumes that
the contingency table size is divisible by the vector unit width. Similar to the
genotype table and contingency table calculations, we can achieve this by
padding the input contingency table with 0's, which will not contribute to the
final MI value. The computation follows the same strategy of avoiding the
calculations of the logarithm of zero as in the regular MI implementation
(Listing~\ref{lst:mi_base}) but by different means: instead of skipping the
logarithm altogether, which is not possible now unless all eight values of the
vector are zero, the function replaces the zeros in the vector registers with
ones that will evaluate to zero and will not contribute in the following FMA
operations (Lines~{\ref{code:vmi256_1}--\ref{code:vmi256_2}},
{\ref{code:vmi256_3}--\ref{code:vmi256_4}} and
{\ref{code:vmi256_5}--\ref{code:vmi256_6}}).

Moving onto \textit{AVX512BW}, this extension provides mask registers and
masked operations, which allow for the execution of operations only on some of
the values contained in the vector register. Masked logarithms are a very
convenient operation to skip the computation of the logarithm of zero. With
masks we can avoid the zeros without having to blend two vector registers
beforehand. Unfortunately, masked logarithms are only available under the SVML
and for a vector width of 512 bits. The rest of the implementations still have
to rely on the sequence of blends and logarithms.

Listing~\ref{lst:mi_avx512f512} shows the same code implemented using 512-bit
intrinsics from the \textit{AVX512BW} extension. Comparisons are now made using
the new intrinsic functions operating with 16-bit masks
(Lines~\ref{code:vmi512_1},~\ref{code:vmi512_2} and~\ref{code:vmi512_3}) instead
of a whole vector register, and the blend operation takes these masks as an
argument. If the SVML is available, the logarithm plus blend sequences of
operations (Lines~\ref{code:vmi512_4},~\ref{code:vmi512_5}
and~\ref{code:vmi512_6}) can be replaced with a single call to the intrinsic
\texttt{\detokenize{mm512_mask_log_ps}}, which only calculates the logarithm on
the positions specified by the mask.

The \textit{AVX512BW} extension also includes mask functions for 256-bit
operations. Therefore, for comparison purposes, a third version of the MI
function using a width of 256 bits was also created. This version uses the same
sequence of blend plus logarithm intrinsic functions shown in
Listing~\ref{lst:mi_avx512f512} both for the SVML and GNU's \textit{libc}
libraries since Intel does not include in the SVML a masked version of the
256-bit logarithm intrinsic.

\subsection{Segmented exploration algorithm}\label{sec:seg_exploration}

Although the algorithm presented in Algorithm~\ref{alg:nonseg} could directly
incorporate the SIMD functions described in the previous subsections, the
performance per core would be penalized due to the interleaved execution of
vector instructions running at very different frequencies. Intel CPUs, such as
the Xeon Gold 6240 used during the evaluation, are known to downscale their CPU
clock frequency based on the number of active cores and the sequence of
instructions executed due to differences in power consumption and/or heat
dissipation. In the processor technical
document~\cite{intel_xeon_specification}, Intel identifies three different
frequency licenses in which the processor operates: non‐AVX, AVX 2.0 and AVX‐512
base core frequencies. Furthermore, different operations inside each license are
not guaranteed to run at the same frequency, these are only base frequencies
that the processor is guaranteed to run at. For example, floating-point
arithmetic vector operations run at a slower clock frequency than integer
arithmetic or bitwise vector operations.

\begin{algorithm}[!t]
  \DontPrintSemicolon{}
  \SetKw{KwOr}{or}
  \KwIn{\;
    \(d\): Data set containing \(n\) genotype tables, each representing a
    single SNP for \(m\) individuals\;
    \(o\): Order of the interactions to identify\;
    \(b\): Size of the block of operations
  }
  \KwOut{\;
    \(v\): Vector of combinations of \(o\) SNPs, and their associated MI
    value
  }
  {\BlankLine}
  {\nl} Allocate an array \(g\) of \(o-2\) genotype tables, for sizes between
    \(1\) and \(o-1\) \;
  {\nl} Allocate an array \(a\) of \(b\) combinations of size \(o\)
    \;\label{code:salg_1}
  {\nl} Allocate an array \(c\) of \(b\) contingency tables of size \(o\)
    \;\label{code:salg_2}
  {\nl} Create an empty stack \(s\) \;
  {\nl} Calculate the marginal entropy \(H(Y)\) \;
  {\nl} \(inv\_inds \leftarrow 1 / m\) \;

  {\nl} \(i \leftarrow 0\) \;
  {\nl} \While{\(s\) is not empty {\KwOr} \(i < n\)}{\label{code:salg_3}
    {\nl} \(l \leftarrow 0\) \;
    {\nl} \While{\(l < b\)}{\label{code:salg_4}
      {\nl} \If{\(s\) is empty}{
        {\nl} \eIf{\(i \ge n\)}{
          {\nl} Break from the inner while loop \;\label{code:salg_5}
        }{
          {\nl} \(g[1] \leftarrow d[i]\) \;\label{code:salg_6}
          {\nl} \For{\(j \leftarrow i + 1\) {\KwTo} \(n\)}{
            {\nl} Push the pair \( \{i,j\} \) into the stack \(s\) \;
          }
          {\nl} \(i \leftarrow i + 1\) \;
          {\nl} Continue on the next iteration of the loop \;\label{code:salg_7}
        }
      }
      {\nl} Pop the combination \( \{c_1,\ldots,c_k\} \) from the top of the
        stack \(s\) \;\label{code:salg_8}
      {\nl} \eIf{\(k < o\)}{
        {\nl} \(g[k] \leftarrow combine(g[k-1], d[c_k])\) \;\label{code:salg_9}
        {\nl} \For{\(j \leftarrow c_k + 1\) {\KwTo} \(n\)}{
          {\nl} Push the new combination \( \{c_1,\ldots,c_k,j\} \) into the
            stack \(s\) \;\label{code:salg_10}
        }
      }{
        {\nl} \(a[n] \leftarrow \{c_1,\ldots,c_k\} \) \;\label{code:salg_11}
        {\nl} \(c[n] \leftarrow combine\_and\_popcnt(\)
          \(g[o-1], d[c_k])\) \;
        {\nl} \(l \leftarrow l + 1\) \;\label{code:salg_12}
      }
    }
    {\nl} \For{\(j \leftarrow 0\) {\KwTo} \(l\)}{\label{code:salg_13}
      {\nl} \(mi\_val \leftarrow MI(c[j], H(Y), inv\_inds)\) \;
      {\nl} Add \(a[j]\) and \textit{mi\_val} to the vector of results
        \textit{v} \;\label{code:salg_14}
    }
  }
  \caption{Segmented exploration}\label{alg:seg}
\end{algorithm}

As a direct consequence of this, the exploration algorithm would run on the
lowest frequency imposed by any of the vector operations, since the change in
frequency is not immediate and depends on the pipeline of operations executed.
To resolve it, Algorithm~\ref{alg:seg} proposes a modification to the algorithm,
segmenting the different operations into blocks of operations corresponding to
similar frequency levels and therefore avoiding the frequency change problem.

Instead of declaring a single contingency table, the function now reserves space
to store \(b\) combinations and compute \(b\) tables before applying MI to any
of them (Lines~\ref{code:salg_1}--\ref{code:salg_2}). Combinations are now
explored using two nested \textit{while} loops, the outer one iterating until
the stack is empty and all combinations starting with the SNP \(i\) have been
explored (Line~\ref{code:salg_3}), and the inner one iterating until the block
of \(b\) contingency tables has been filled (Line~\ref{code:salg_4}).

Every iteration of the innermost loop starts by checking if the stack is empty.
If that is the case, and there are no more SNPs to explore, the loop exits
(Line~\ref{code:salg_5}); otherwise, the genotype table of the SNP \(i\) is
copied in \(g[1]\), all combinations of two SNPs starting with \(i\) are pushed
into the stack, the counter \(i\) is increased and the execution continues on
the next iteration of the inner while loop
(Lines~\ref{code:salg_6}--\ref{code:salg_7}). If the stack is not empty, the
loop operates in a similar manner as the old one: the top combination \(
\{c_1,\ldots,c_k\} \) of the stack is popped (Line~\ref{code:salg_8}). If \(k\)
is smaller than the target interaction order \(o\), its genotype table is
computed and stored in the array of genotype tables \(g\), and all subsequent
combinations starting with \( \{c_1,\ldots,c_k\} \) are pushed into the stack
(Lines~\ref{code:salg_9}--\ref{code:salg_10}). Otherwise, \( \{c_1,\ldots,c_k\}
\) and its contingency table are stored in the \(a\) and \(c\) arrays,
respectively (Lines~\ref{code:salg_11}--\ref{code:salg_12}). When the arrays
\(a\) and \(c\) of \(b\) index combinations and tables, respectively, has been
filled, the inner \textit{while} finishes and a \textit{for} loop iterates over
all the computed contingency tables, calculating its MI and adding the
combination into the vector of results \(v\)
(Lines~\ref{code:salg_13}--\ref{code:salg_14}).

The selection of a proper value for the block size \(b\) is key in order to
obtain good performance. It has to be large enough to make the impact of the
transition between frequencies negligible, but not large enough to exceed the
second-level cache of the processor. Through experimental testing, we found that
an appropriate \(b\) for the Intel Xeon Gold 6240, the processor used in the
evaluation, is \(1474560 / 3^o\), with \(o\) being the order of the search.
This size corresponds to the smallest block size tested at which the average
running frequencies of the functions is very close or equal to the running
frequency of these same functions in isolation.

\section{Evaluation}\label{sec:evaluation}

We have conducted an extensive evaluation of the performance achieved by the
automatic vectorization offered by the GCC and Intel compilers, in contrast to
manual vectorization using Intel Intrinsics, when implementing a SIMD epistasis
detection algorithm. It considers the performance of the different functions
that compose the epistasis search in isolation, as well as the whole depth-first
search algorithm. This evaluation starts by assaying the individual functions
separately, and identifying which of the implementations obtains the best
performance in each part. Then, the search algorithm is evaluated showcasing how
the relative differences in time spent in each of the functions, and the
operations that each function involves, influence the performance of the whole
search. At last, the best performing implementation is compared against the
original MPI3SNP~\cite{ponte-fernandez_fast_2019} program using the compiler's
automatic vectorization, to put into perspective the performance gain achieved.

Given the exponential time complexity of the operations that compose an
epistasis search, and the search itself, it is difficult to represent elapsed
time results for different problem sizes in the same graph and extract
conclusions from them. For this reason, this evaluation uses the average elapsed
time per cell or row (depending on the computation being evaluated) as the
metrics to present the results. These measures of time express the compute time
relative to the complexity of the computation, thus removing the impact of this
complexity from the results and highlighting the differences in performance from
multiple implementations of the same operation.

The two compilers used throughout the evaluation are the GNU C Compiler 8.3.0
(with the GNU \textit{libc} version 2.29) and the Intel C++ Compiler 2020
(version 19.1.1.217). The same optimization flags were used for both compilers:
\texttt{-O3}, \texttt{-march={\allowbreak}native} and \texttt{-mtune=native}.
Additionally, we enabled optimizations on floating-point arithmetic operations
using \texttt{-fast-{\allowbreak}math} and \texttt{-fp-model=fast} for the the
two compilers respectively, as it is a requirement for GCC in order to vectorize
some calls to the math library. Furthermore, for the automatic vectorization, we
considered the effects of indicating a preference for a particular vector width
during the compilation through the flags
\texttt{-mprefer-vector-width=\{256,512\}} for the GCC compiler and
\texttt{-qopt-zmm-usage=\{low,high\}} for the Intel compiler. Only the flag
\texttt{-qopt-zmm-usage=high} had a positive impact on performance, thus it is
the only one included in the results.

All experiments were run on an Intel Xeon Gold 6240, an 18-core CPU that
implements the \textit{AVX2}, \textit{AVX512F}, \textit{AVX512CD},
\textit{AVX512BW}, \textit{AVX512DQ}, \textit{AVX512VL} and \textit{AVX512VNNI}
vector extensions. As mentioned in Section~\ref{sec:seg_exploration},
performance during SIMD operation in modern Intel CPUs is tied to the number of
active threads and the type of vector operations used in the instruction
pipeline. This is stated in the processor technical
document~\cite{intel_xeon_specification}, where the maximum core frequencies in
turbo mode are specified attending to the number of cores and type of vector
operation used. Therefore, to obtain a realistic multithreaded performance,
elapsed times throughout the evaluation are measured during a simultaneous
execution of the function in question on every core of the processor. The 18
different times are then averaged and presented as a single value.

\begin{figure*}[ht]
  \centering
  \subfloat[GCC compiler]{%
    \includegraphics[width=.49\textwidth]{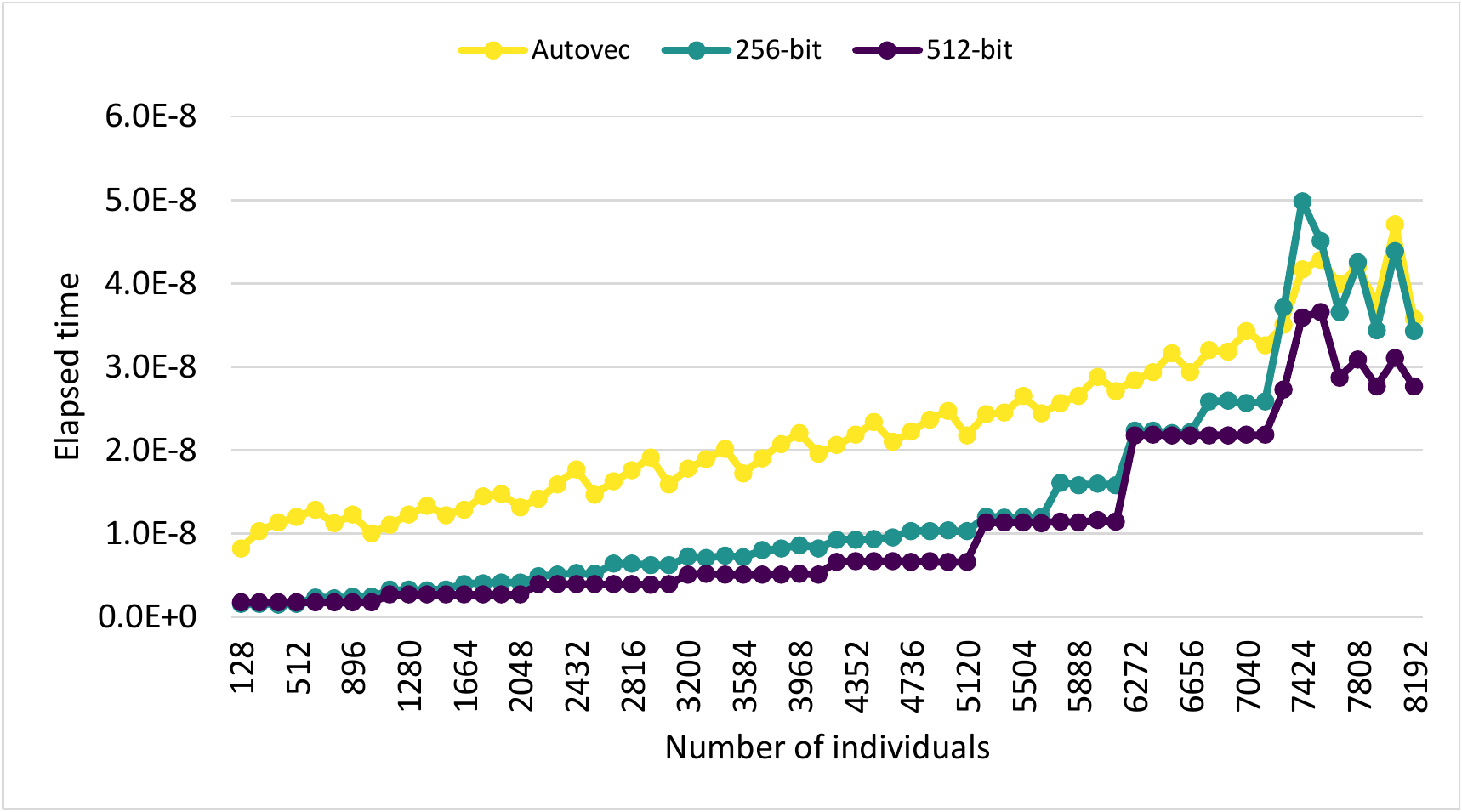}
  }
  \subfloat[Intel C++ compiler]{%
    \includegraphics[width=.49\textwidth]{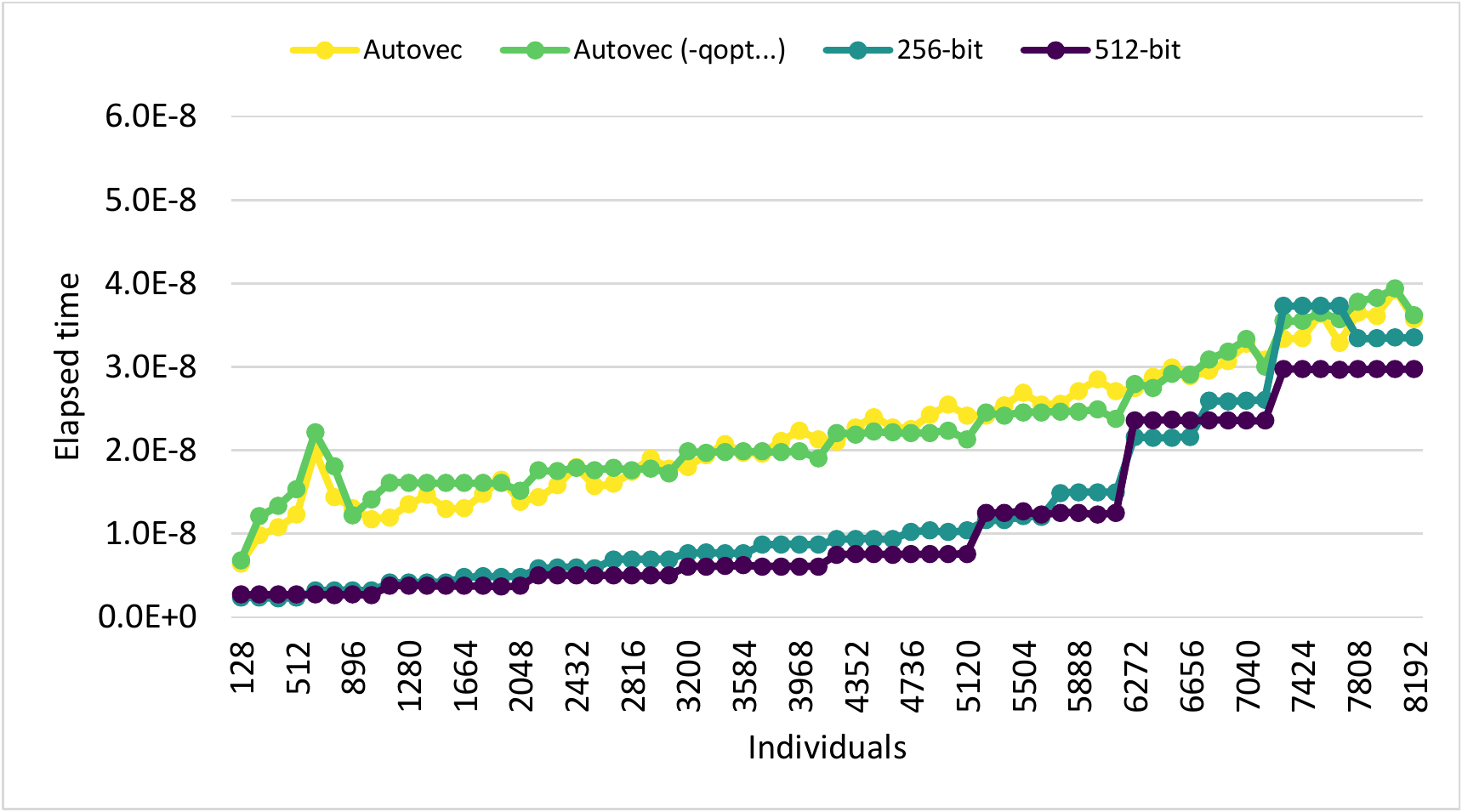}
  }
  \caption{
    Average elapsed time per row during the calculation of genotype tables, for
    an increasing number of individuals and a fixed combination size of three,
    both using the GNU C Compiler and the Intel C++ Compiler.
  }\label{fig:gt_individuals}
\end{figure*}

\subsection{Genotype table calculation performance}

\begin{figure*}[ht]
  \centering
  \subfloat[GCC compiler]{%
    \includegraphics[width=.49\textwidth]{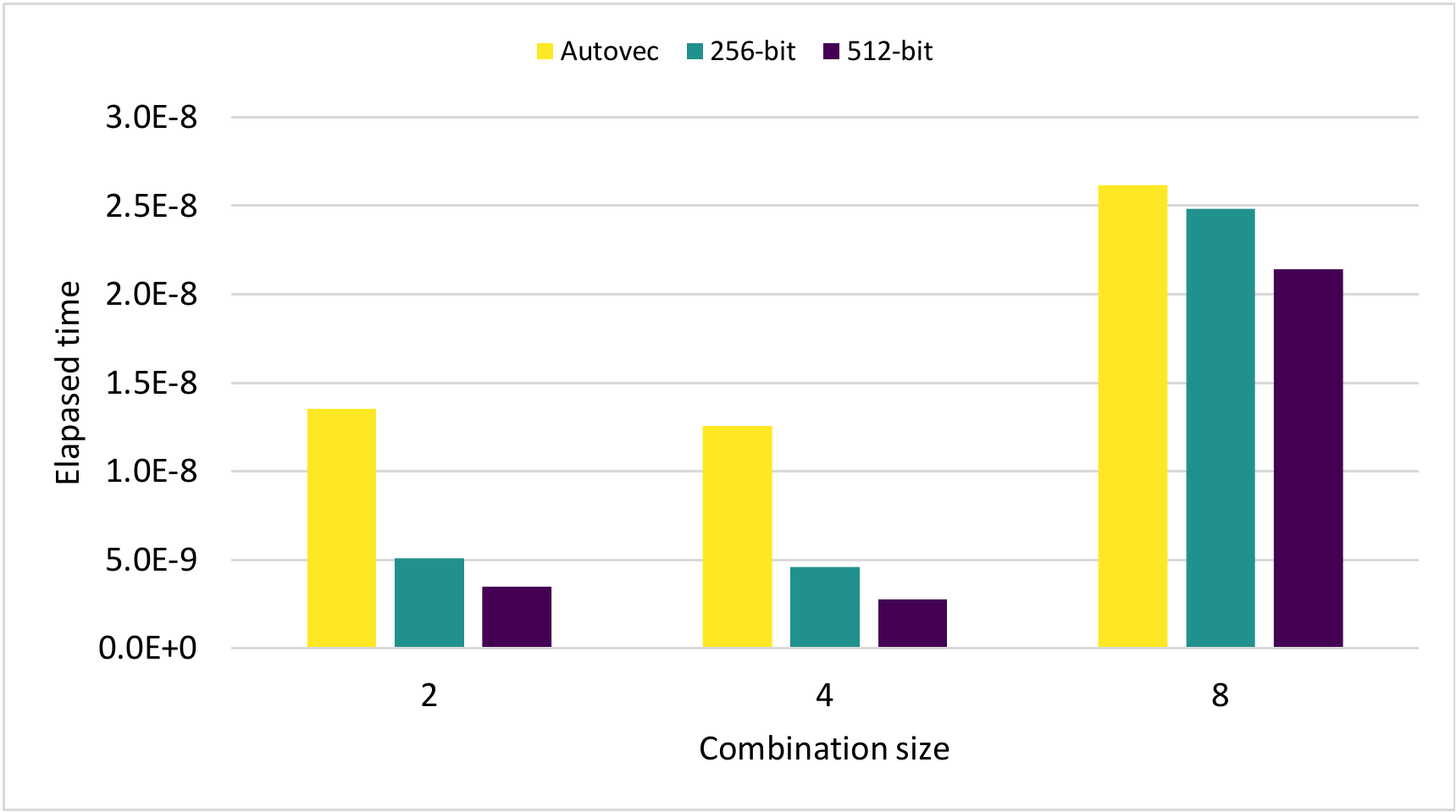}
  }
  \subfloat[Intel C++ compiler]{%
    \includegraphics[width=.49\textwidth]{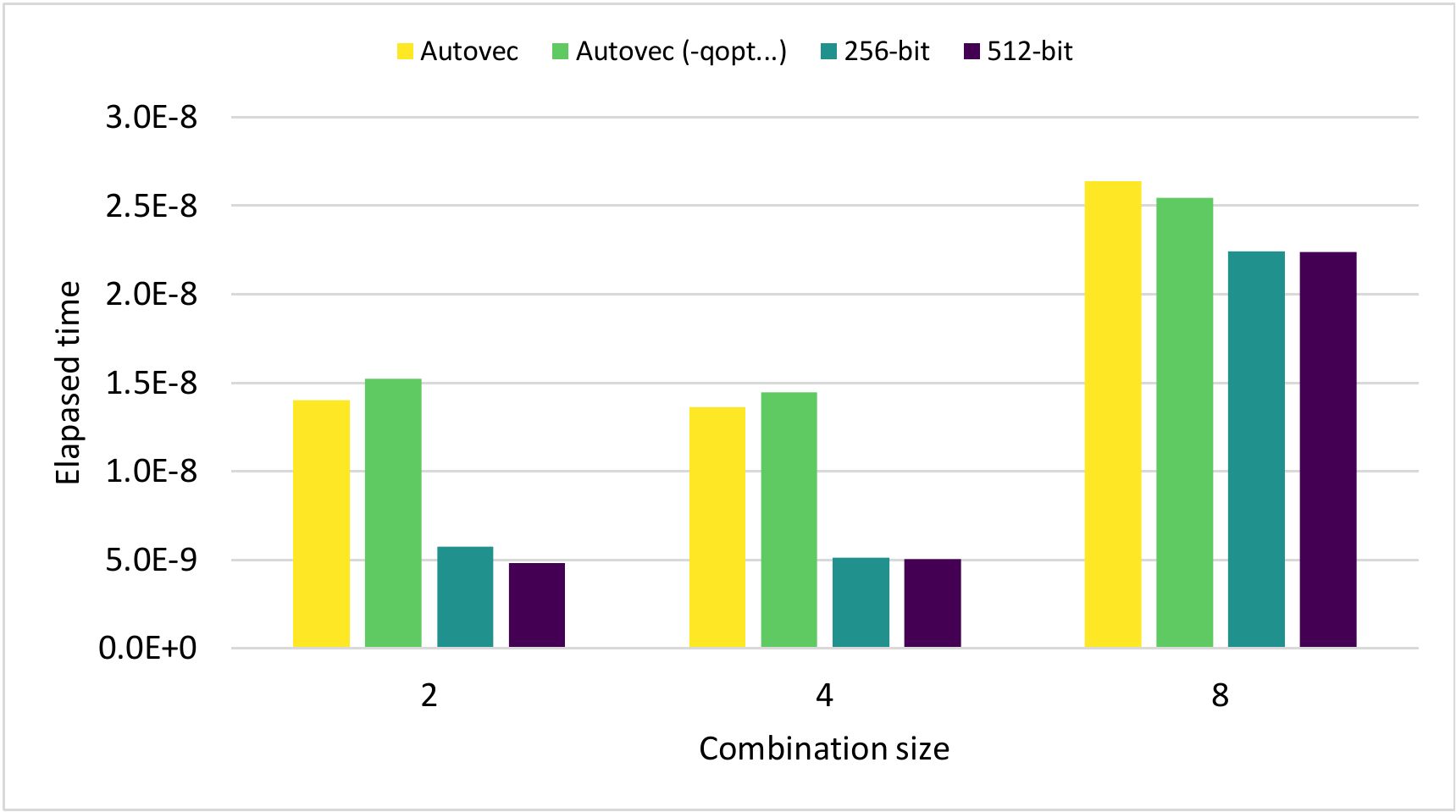}
  }
  \caption{
    Average elapsed time per row during the calculation of genotype tables, for
    combination sizes of 2, 4 and 8 and a fixed number of 2048 individuals, both
    using the GNU C Compiler and the Intel C++ Compiler.
  }\label{fig:gt_order}
\end{figure*}

Figs.~\ref{fig:gt_individuals}~and~\ref{fig:gt_order} represent the performance
results for the genotype table calculation function. The figures compare the
performance of the explicit vectorization using 256-bit and 512-bit vector
instructions with the automatically vectorized code, both for the GCC and Intel
compilers. The measure of time used in both figures is the average time per row,
that is, the average elapsed time during the calculation of a single row of the
table including both cases and controls, for all of the genotype tables of the
order and number of individuals specified.

Both compilers are capable of automatically vectorizing this function with no
problems. Despite this, and as the figures show, the performance of the
autovectorization for both compilers is inferior than the performance of both
explicitly vectorized alternatives.

Fig.~\ref{fig:gt_individuals} represents the time per row during the computation
of genotype tables corresponding to a combination of three SNPs, for a growing
number of individuals. The time per row grows linearly with the number of
individuals, as every row of the genotype table contains information about all
the individuals in the data. Both compilers show a gap between the performance
of the automatic and explicit vectorizations that is present until a number of
individuals higher than 7040. The 512-bit explicit implementation performs
slightly better in general than the 256-bit one.

Fig.~\ref{fig:gt_order} represents the time per row during the computation of
genotype tables corresponding to combinations of 2, 4 and 8 SNPs, using a fixed
number of individuals of 2048. Here, the time per row should remain constant
when increasing the size of the combinations (\(s\)), as the number of rows in
the table (\(3^s\)) grows with the number of SNPs in combination considered, but
each row contains the same 2048 individuals. This is the case for combination
sizes smaller than eight, where the time per row remains mostly constant between
combination sizes of two to seven. Starting at genotype tables of eight SNPs,
there is an increase in the elapsed time due to the size of the operands and
result tables exceeding the level 1 data cache of the processor, which is
manifested in the results by bringing the vector and non-vector performances
much closer.

For second and fourth-order interactions, both explicit vectorization
alternatives obtain again better results than the vectorization applied by the
compiler, with the 512-bit implementation performing the best. For eighth order
interactions the performance gap is smaller, with less relative difference
between implementations.

\begin{figure*}[!ht]
  \centering
  \subfloat[GCC compiler]{%
    \includegraphics[width=.49\textwidth]{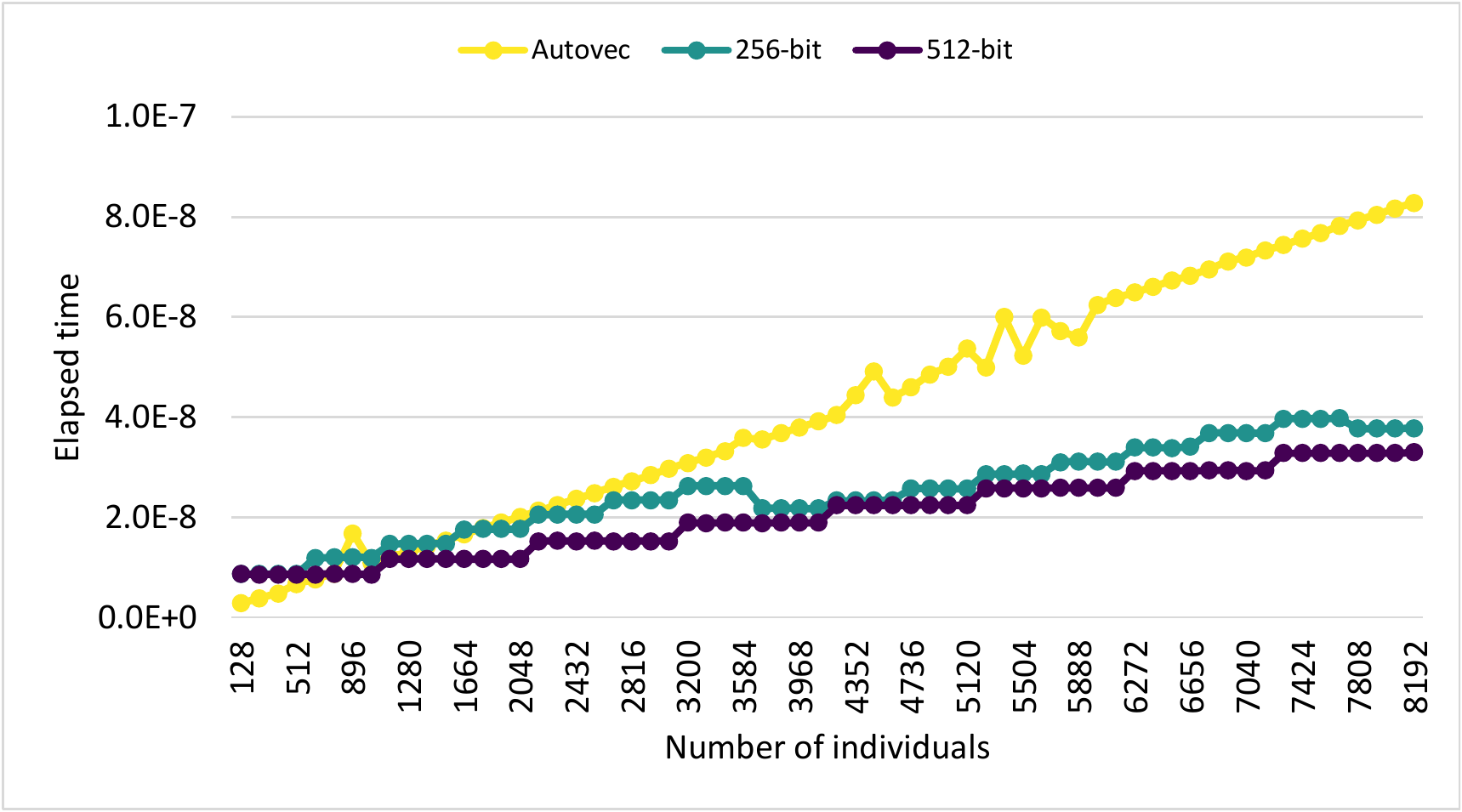}
  }
  \subfloat[Intel C++ compiler]{%
    \includegraphics[width=.49\textwidth]{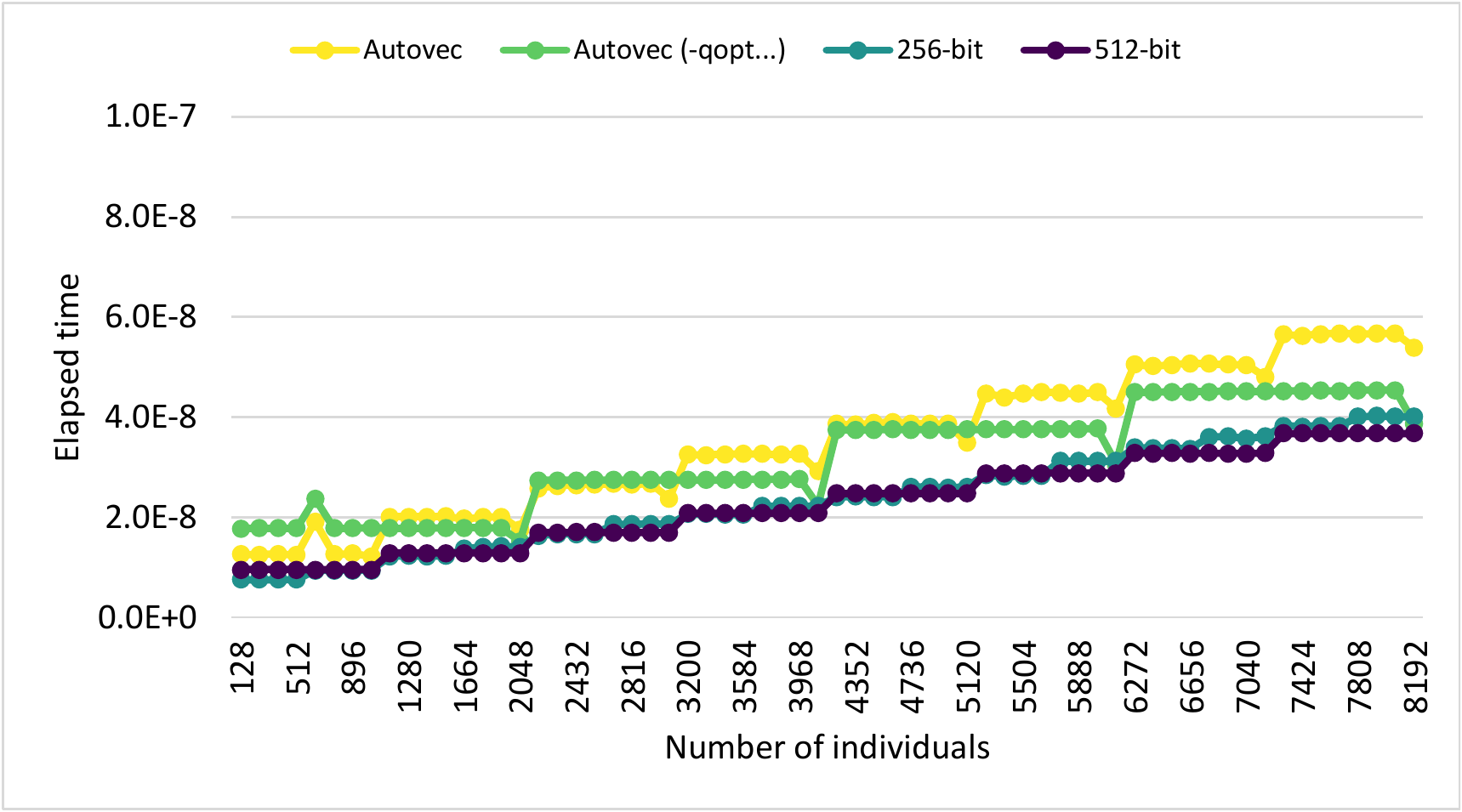}
  }
  \caption{
    Average elapsed time per cell during the calculation of contingency tables,
    for an increasing number of individuals and a fixed combination size of
    three, both using the GNU C Compiler and the Intel C++ Compiler.
  }\label{fig:ct_individuals}
\end{figure*}

When taking a look at the frequencies at which the different implementations
run, we observe that the genotype table calculation runs at 3270 MHz for the
256-bit vector width and 2805 MHz for the 512-bit vector width. In a different
architecture, or in future Intel microarchitectures, where the difference in
frequencies between vector widths could be smaller or nonexistent, we can expect
the performance gap between the two widths to be larger. As an example, the
elapsed time per row of calculating a fourth-order genotype table of 2048
individuals at a fixed frequency of 2.6 GHz (the base frequency of the
processor) is \num{6.52e-09}s and \num{3.00e-09}s for the explicit 256-bit and
512-bit implementations under GCC, respectively; and \num{6.32e-09}s and
\num{4.04e-09}s for the same implementations under Intel, respectively. That is,
the 512-bit implementation is 2.18 and 1.56 times faster than the 256-bit one
for each compiler, significantly larger than what we observe in
Fig.~\ref{fig:gt_order} between these two implementations (1.66 and 1.01).

\subsection{Contingency table calculation performance}

\begin{figure*}[!ht]
  \centering
  \subfloat[GCC compiler]{%
    \includegraphics[width=.49\textwidth]{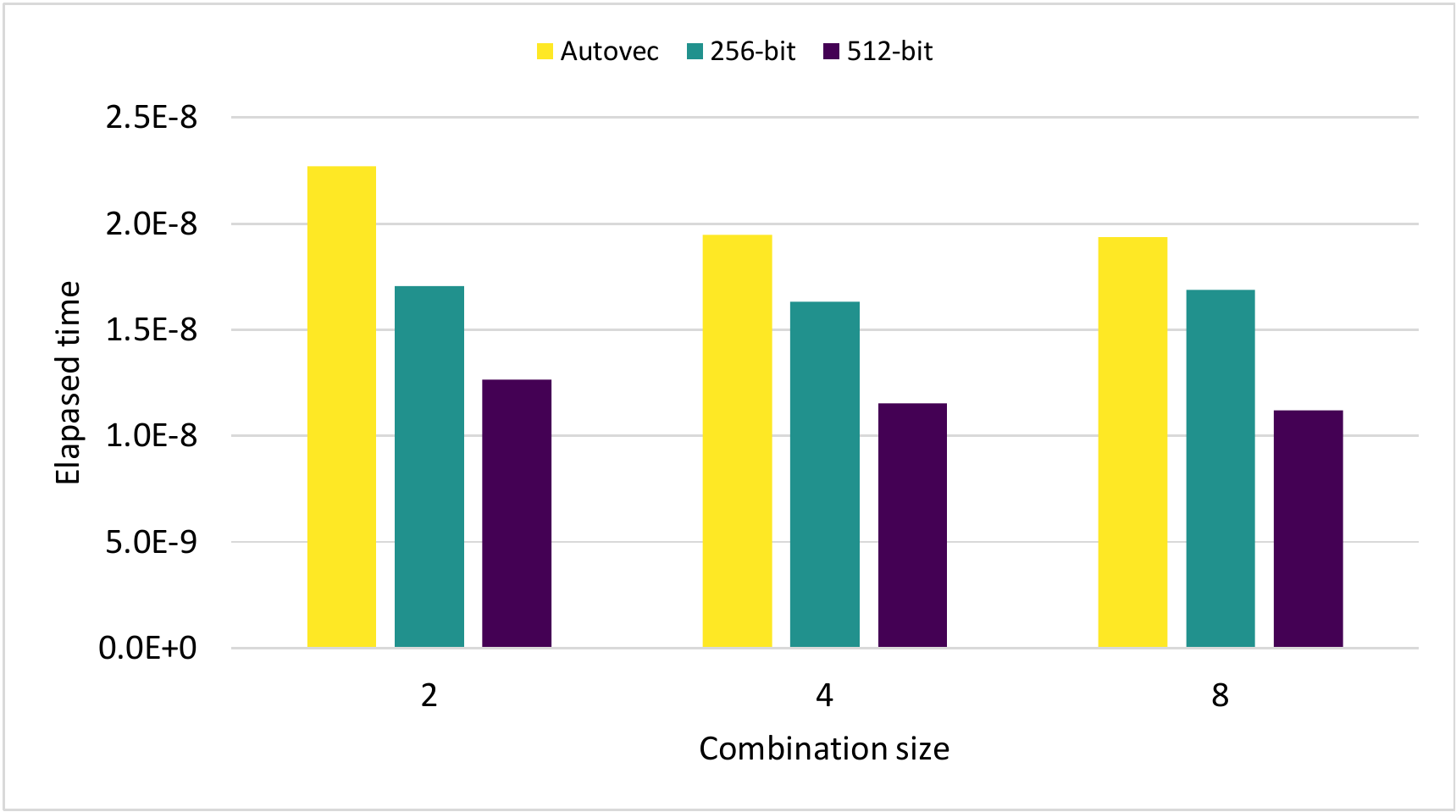}
  }
  \subfloat[Intel C++ compiler]{%
    \includegraphics[width=.49\textwidth]{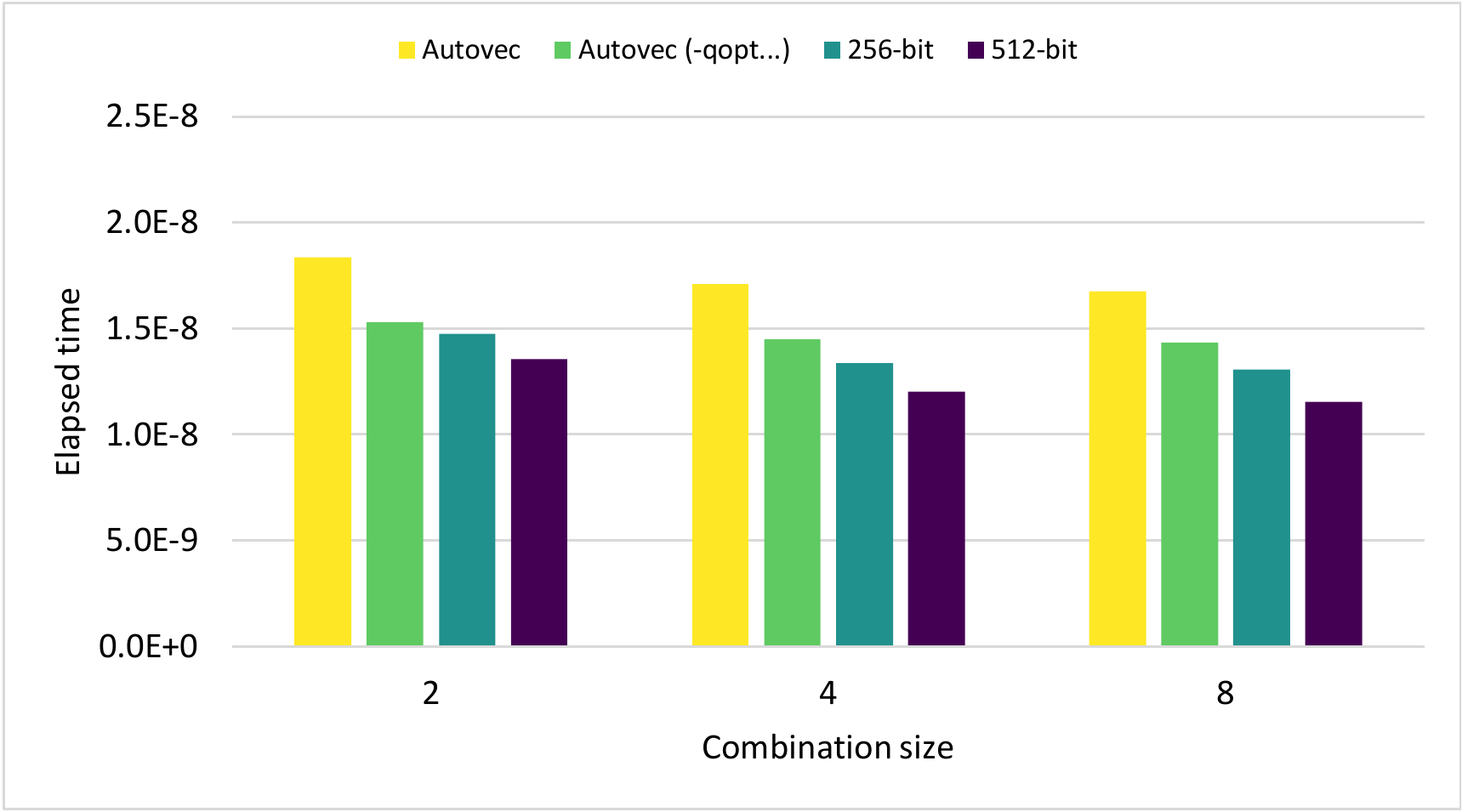}
  }
  \caption{
    Average elapsed time per cell during the calculation of contingency tables,
    for combination sizes of 2, 4 and 8 and a fixed number of 2048 individuals,
    both using the GNU C Compiler and the Intel C++ Compiler.
  }\label{fig:ct_order}
\end{figure*}

Figs.~\ref{fig:ct_individuals}~and~\ref{fig:ct_order} represent the performance
results for the contingency table calculation function. Similar to the figures
from the genotype table calculation, these also compare the performance of the
explicit vectorization using 256-bit and 512-bit vector operations with the
automatically vectorized code using both GCC and Intel compilers. In this case,
we use the average elapsed time per cell to represent the performance results,
that is, the average time for the calculation of a single cell of the table, for
all of the contingency tables of the order and number of individuals specified.

For this function, only the Intel compiler is capable of vectorizing the
\textit{popcount} operation via the introduction of its own vector
implementation. GCC, on the other hand, refuses to vectorize this function due
to the presence of the aforementioned operation inside the innermost loop.

Fig.~\ref{fig:ct_individuals} represents the time per cell during the
computation of contingency tables corresponding to a combination of three SNPs,
for a growing number of individuals. The elapsed time per cell during the
creation of contingency tables also grows linearly with the number of
individuals, since the function operates with the rows from the two previous
genotype tables, which include the data of all individuals. The differences in
compiler behaviour are apparent: GCC results display a linear increase of the
elapsed time per cell with the number of individuals, at a faster pace than the
Intel results due to the lack of vectorization.  It is also worth noting that
there is a small reduction of the elapsed time per cell in the explicit 256-bit
vectorization under GCC at 3712 individuals, which corresponds to the minimum
number of individuals required to enter the loop that includes unrolling
(Listing~\ref{lst:ct_avx2}, Line~\ref{code:vct256_3}). Anyhow, both explicit
implementations are faster than the codes that the two compilers offer, with the
512-bit explicit vectorization being the fastest alternative.

Fig.~\ref{fig:ct_order} represents the time per cell during the computation of
contingency tables for combination sizes of 2, 4 and 8, and using the same
number of 2048 individuals. Analogous to the elapsed time per row during the
calculation of genotype tables, the time per cell should also remain constant
with the size of the combinations explored. However, contrary to those results,
there is no increase in the elapsed time for eight order tables due to cache
problems thanks to the avoidance of the genotype table storage. This is due to
the merge of the last level genotype table calculations and contingency table
computations in a single function. Results show that the explicit
implementations are faster than the compiler-generated code, with the 512-bit
vectorization being the fastest implementation.

\begin{figure*}[t]
  \centering
  \subfloat[GCC compiler]{%
    \includegraphics[width=.49\textwidth]{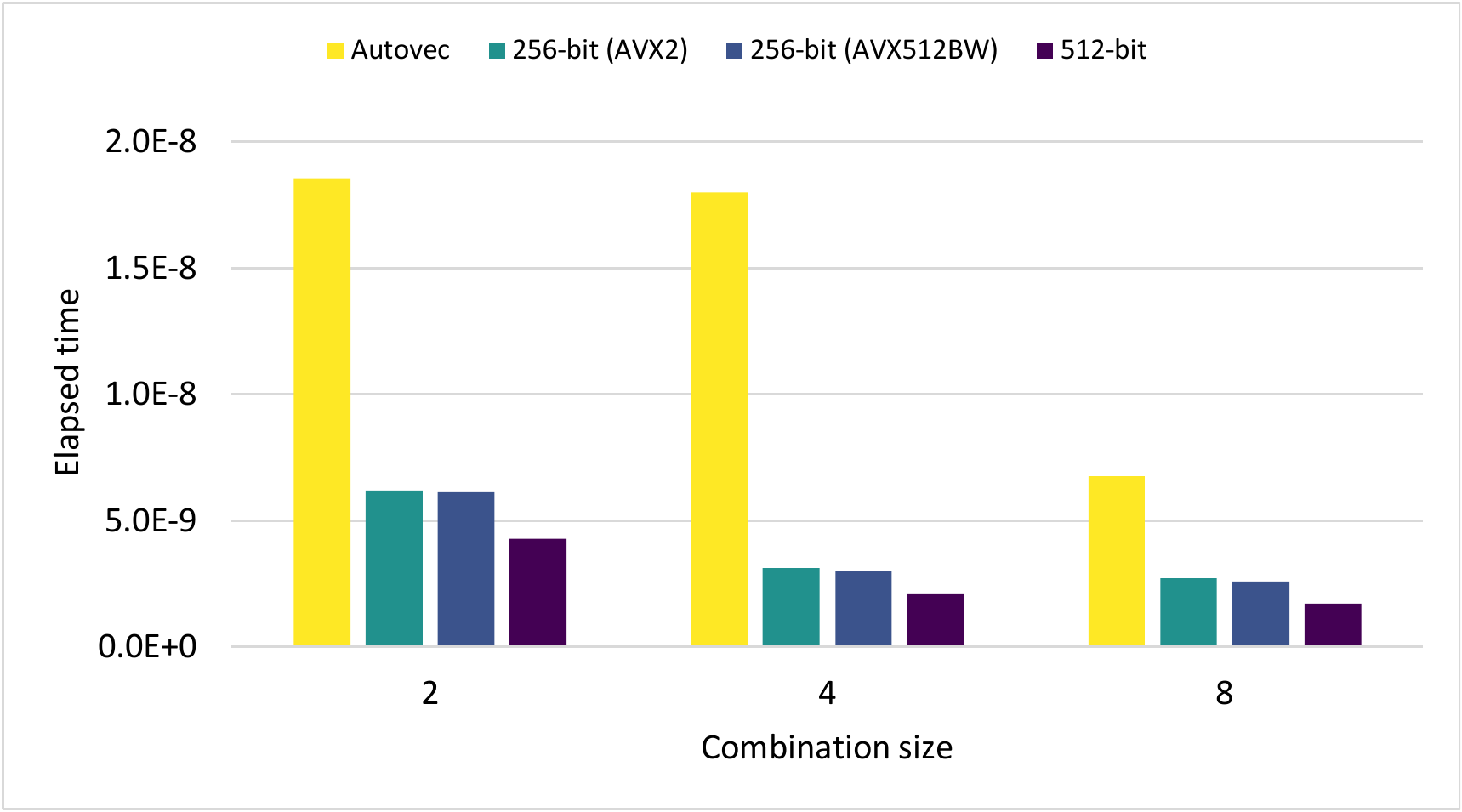}
  }
  \subfloat[Intel C++ compiler]{%
    \includegraphics[width=.49\textwidth]{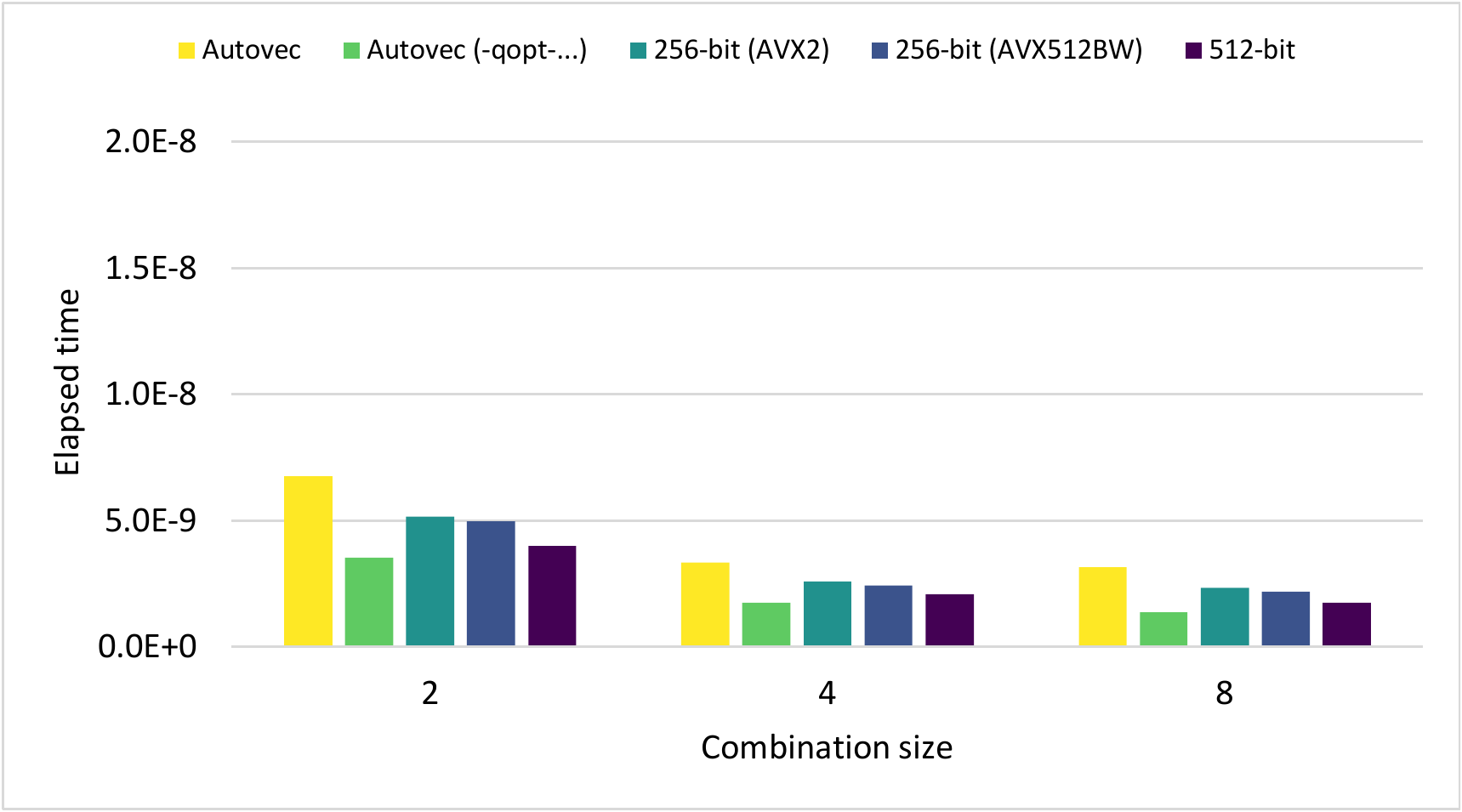}
  }
  \caption{
    Average elapsed time per cell during the calculation of the MI metric, for
    combination sizes of 2, 4 and 8, both using the GNU C Compiler and the Intel
    C++ Compiler.
  }\label{fig:mi}
\end{figure*}

Similar to the genotype table calculations, we observe that the contingency
table computation function runs at a frequency of 3195 MHz for the 256-bit
vector width and 2800 MHz for the 512-bit vector width. If we run the function
at a fixed frequency of 2.6 GHz (the base frequency of the processor), the
elapsed time per cell of calculating a fourth-order contingency table of 2048
individuals is \num{1.97e-08}s and \num{1.24e-08}s for the explicit 256-bit and
512-bit implementations under GCC, respectively; and \num{1.58e-08}s and
\num{1.29e-08}s for the same implementations under Intel, respectively. This
represents a speedup of 1.59 and 1.23 between vector widths for each compiler,
slightly larger than those observed in Fig.~\ref{fig:ct_order} (1.42 and 1.11).

\subsection{Mutual Information calculation performance}

Fig.~\ref{fig:mi} shows the performance results for the MI computation function.
This figure compares the performance of the automatically vectorized code with
three explicit implementations: two 256-bit vector implementations using
Intrinsics from the \textit{AVX2} and \textit{AVX512BW} extensions,
respectively, and a  512-bit vector implementation using Intrinsics from the
\textit{AVX512BW} extension. Here we also use the time per cell to represent the
performance results. This time measures the average elapsed time during the MI
calculations corresponding to a single cell of the contingency table (both for
cases and controls), for all of the tables of the specified order. The number of
cells of a contingency table only depends on the number of SNPs in combination.
Thus, MI, as opposed to the previous routines, does not depend on the number of
individuals.

Results show that the time per cell for the explicit vector implementations
generally decreases with the table size, despite the fact that, ideally, the
workload per contingency table cell should remain constant regardless of the
size of the table. This can mostly be attributed to the additional computations
derived from the padding introduced in the input contingency tables. The larger
the tables are, the lower the number of unnecessary computed cells is relative
to the total number of cells in the table.

The best vector performance is achieved by the automatic vectorization of the
Intel Compiler when coupled with the flag \texttt{-qopt-zmm-usage=high}. In
contrast, GCC's automatic vectorization does not vectorize the loop, despite
having a vectorized logarithm function available in the GNU \textit{libc} math
library. Explicit vectorization using 512-bit AVX instructions obtains the best
performance out of the explicit vector implementations (for GCC it is the
fastest alternative). Furthermore, the introduction of 256-bit \textit{AVX512BW}
instructions in the function have no significant impact on the elapsed time when
compared to the \textit{AVX2} implementation.

When examining the assembly code to characterize the difference in performance
between the explicit vector implementations and the code that the Intel
auto-vectorizer generates, we found that the Intel Compiler calls a function
from the SVML that is not available using Intrinsics:
\texttt{\detokenize{__svml_logf8_mask_e9}} (a logarithm function for a vector
width of 256 bits that uses a masked input). Therefore, in some scenarios,
explicit vectorization may never obtain a performance equal or better than
Intel's auto-vectorization due to the difference in SVML functions available
through Intrinsics.

As for the frequencies at which the function is executed, the 256-bit
implementation runs at 2805 MHz (we only measured the 256-bit implementation
using \textit{AVX2} Intrinsics, since the elapsed times are almost the same)
while the 512-bit implementation runs at 2500 MHz. These frequencies are
considerably lower than the two previous functions due to the usage of
floating-point arithmetic operations. If the running frequencies are fixed to
2.5 GHz (slightly lower than the base frequency because the 512-bit
implementation runs at this frequency), we observe that the elapsed time per
cell of calculating the MI of a fourth-order contingency table are
\num{3.48e-09}s and \num{2.10e-09}s for the 256-bit and 512-bit vectorizations
under GCC, respectively, and \num{2.87E-09}s and \num{2.10E-09}s under the Intel
Compiler. This represents a speedup of 1.66 and 1.36 between vector widths for
each compiler, respectively, slightly larger than those observed in
Fig.~\ref{fig:mi} (1.49 and 1.23).

\subsection{Exhaustive search performance}

\begin{figure*}[tp]
  \centering
  \subfloat[GCC compiler, without segmentation]{%
    \includegraphics[width=.4\textwidth]{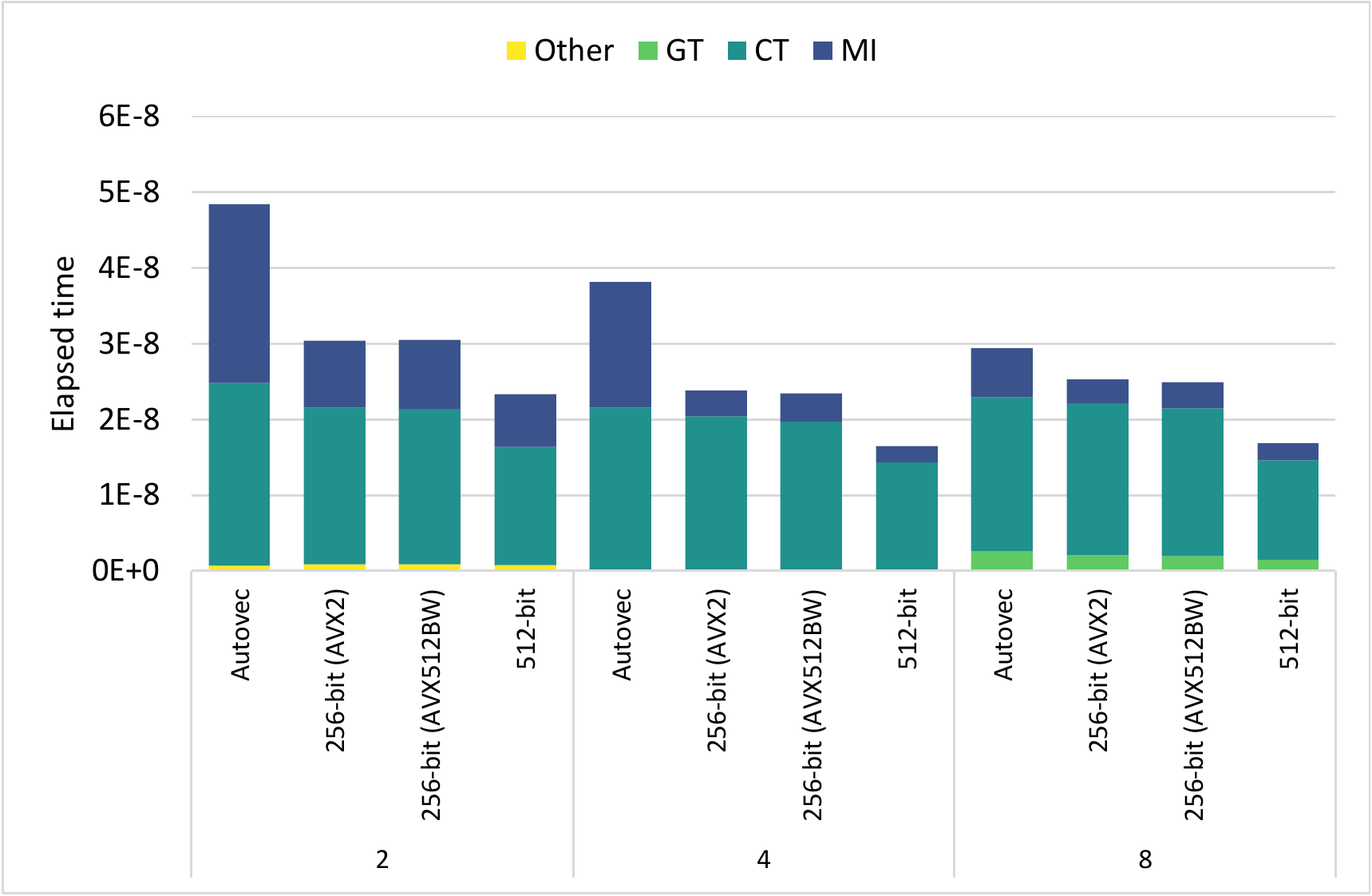}
  }
  \subfloat[GCC compiler, with segmentation]{%
    \includegraphics[width=.4\textwidth]{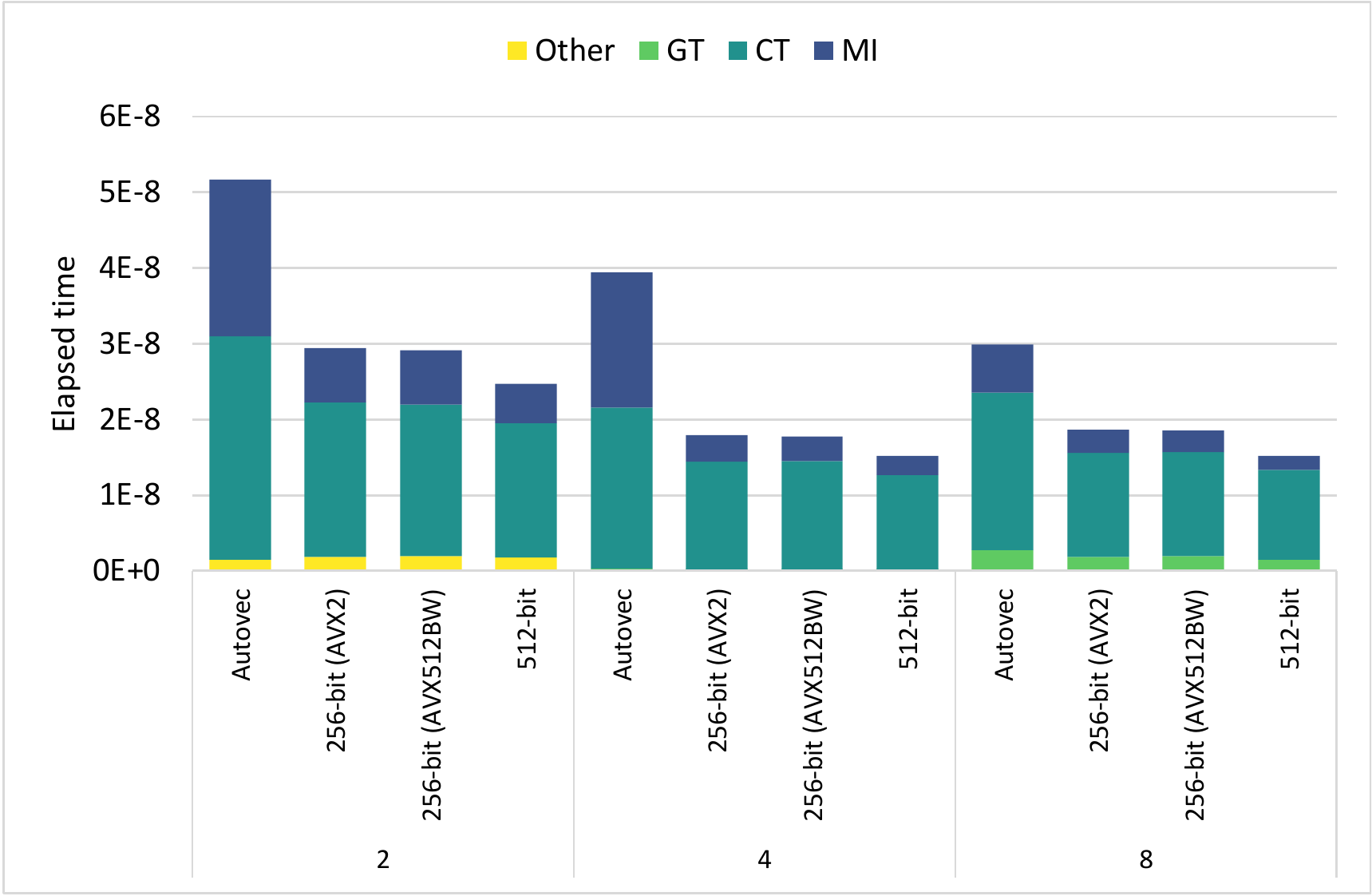}
  }
  \\
  \subfloat[Intel compiler, without segmentation]{%
    \includegraphics[width=.4\textwidth]{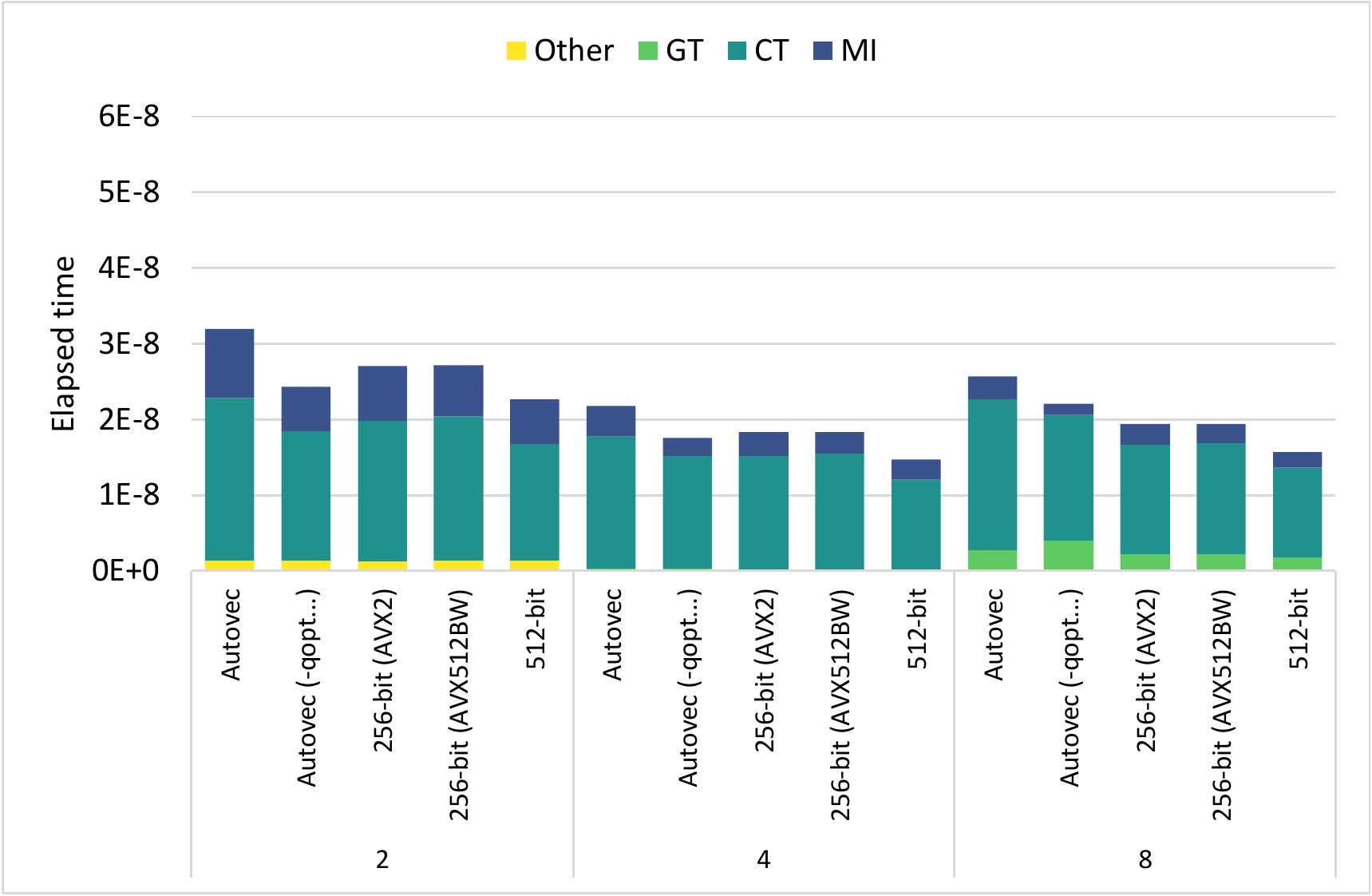}
  }
  \subfloat[Intel compiler, with segmentation]{%
    \includegraphics[width=.4\textwidth]{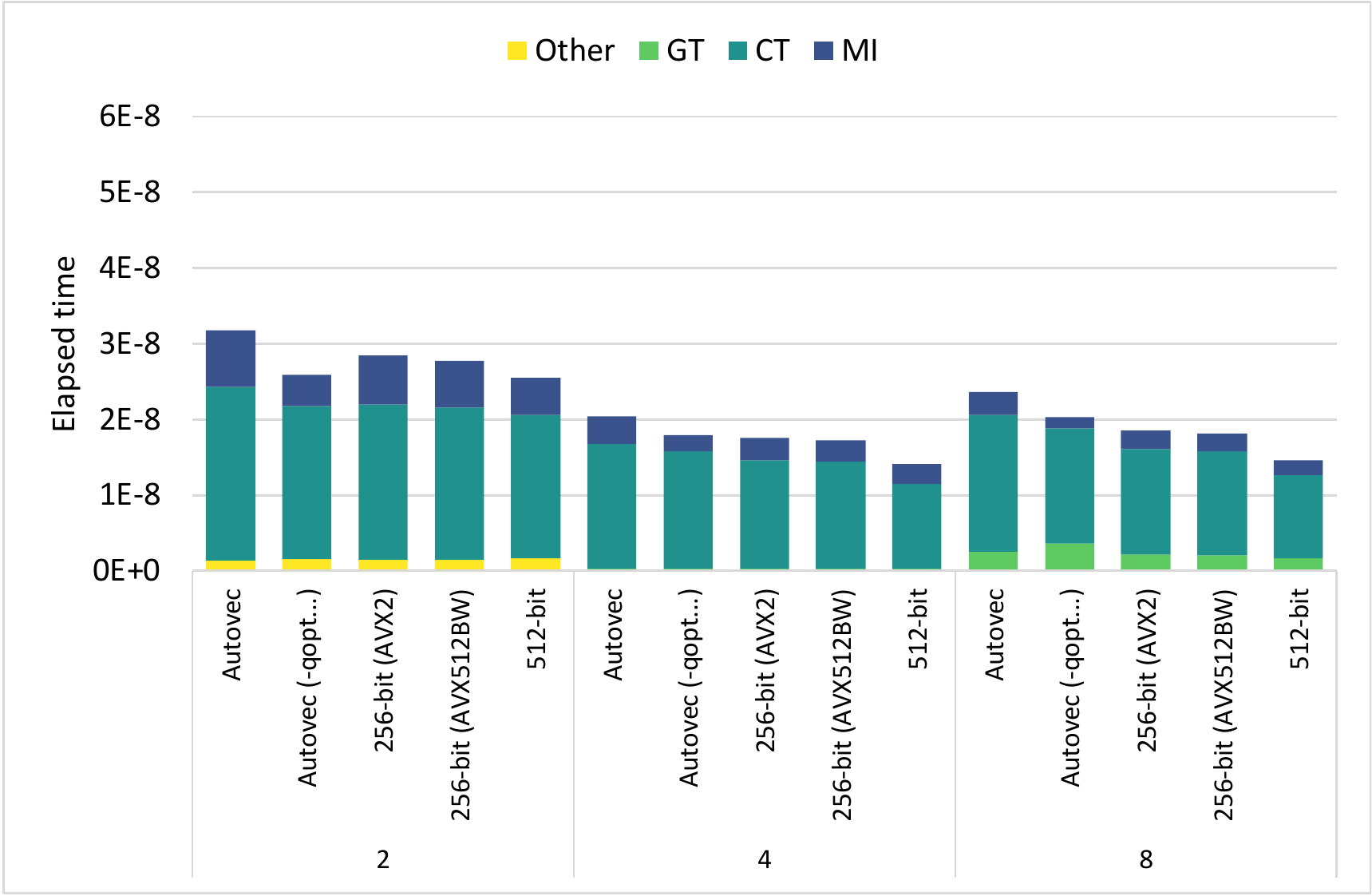}
  }
  \caption{
    Average elapsed time per cell during the exhaustive search of epistasis, for
    combination sizes of 2, 4 and 8 and a fixed number of 2048 individuals, both
    using the GCC and Intel compilers. The times for each approach is divided
    into the calculation of the Genotype Tables (GT), Contingency Tables (CT),
    Mutual Information (MI) and the rest of the operations included in the
    algorithm.
  }\label{fig:search_order}
\end{figure*}

\begin{figure*}[tp]
  \centering
  \subfloat[GCC compiler, without segmentation]{%
    \includegraphics[width=.4\textwidth]{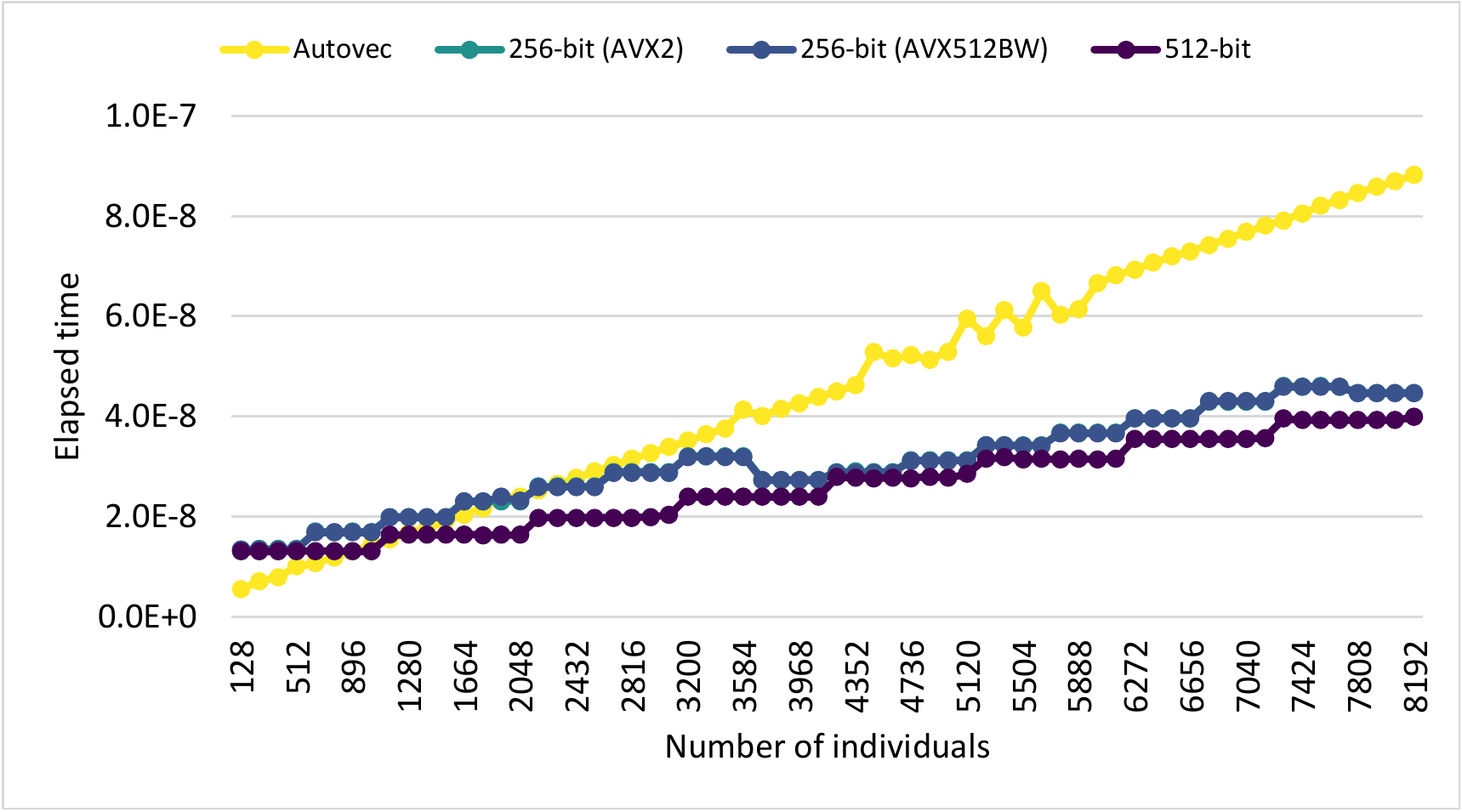}
  }
  \subfloat[GCC compiler, with segmentation]{%
    \includegraphics[width=.4\textwidth]{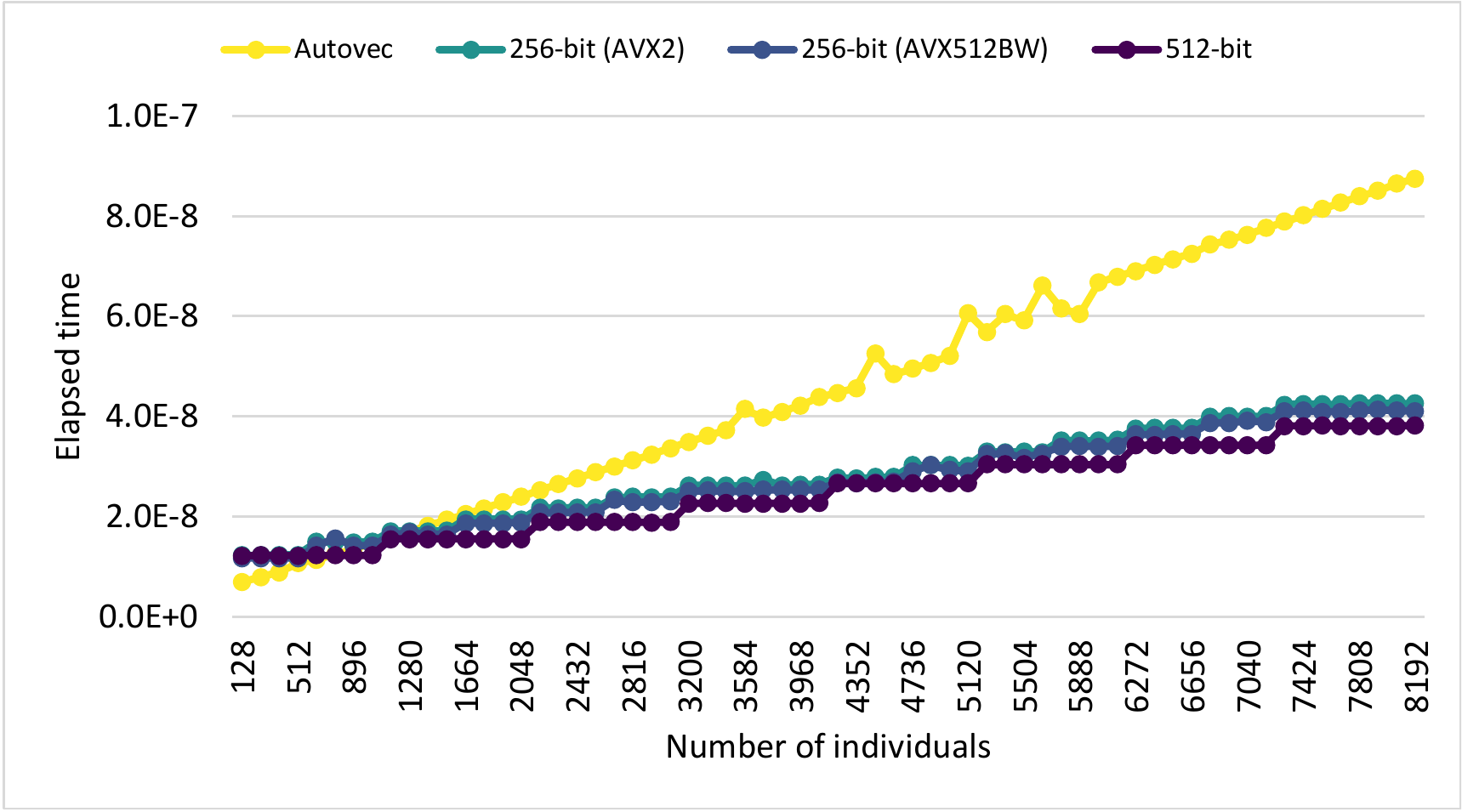}
  }
  \\
  \subfloat[Intel compiler, without segmentation]{%
    \includegraphics[width=.4\textwidth]{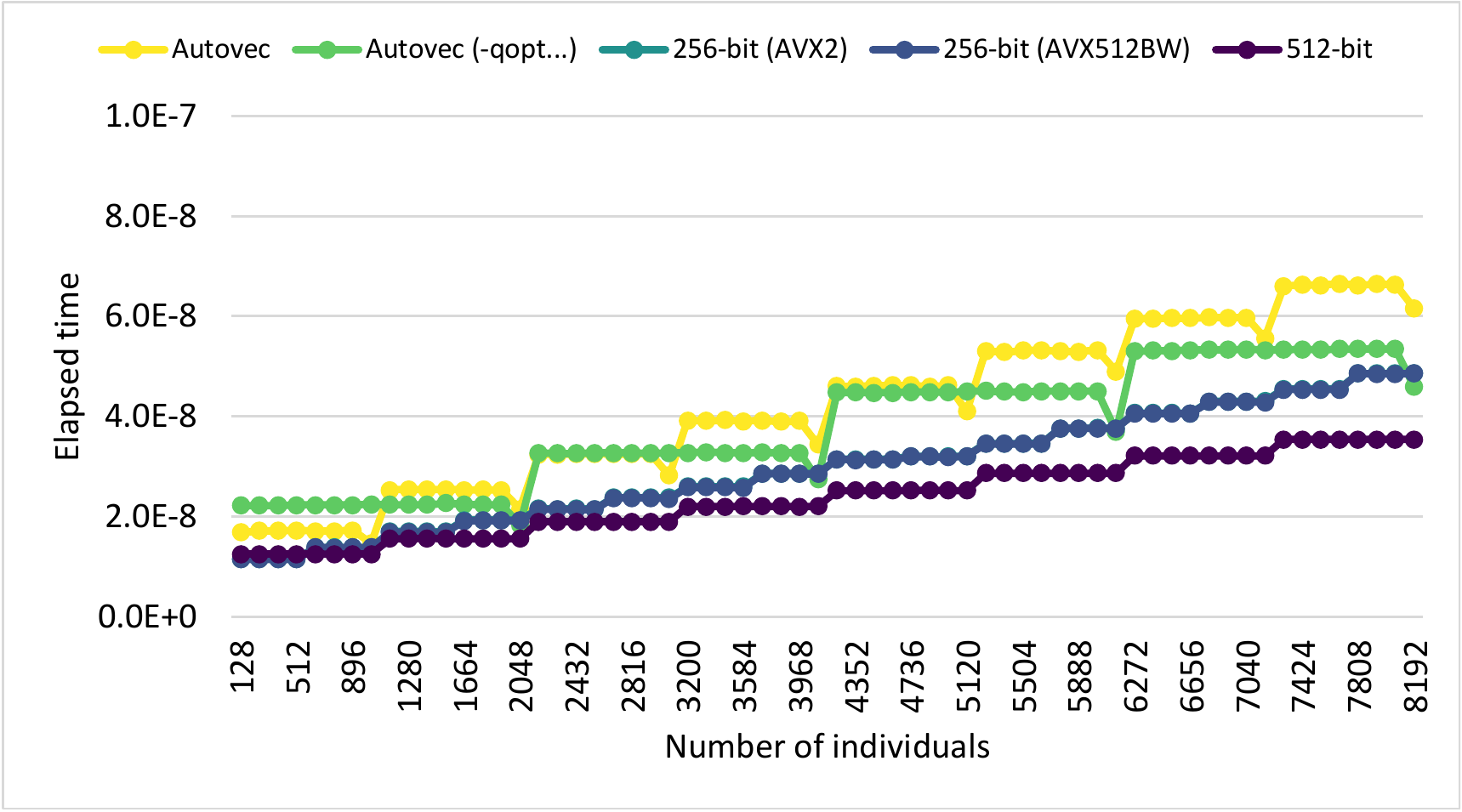}
  }
  \subfloat[Intel compiler, with segmentation]{%
    \includegraphics[width=.4\textwidth]{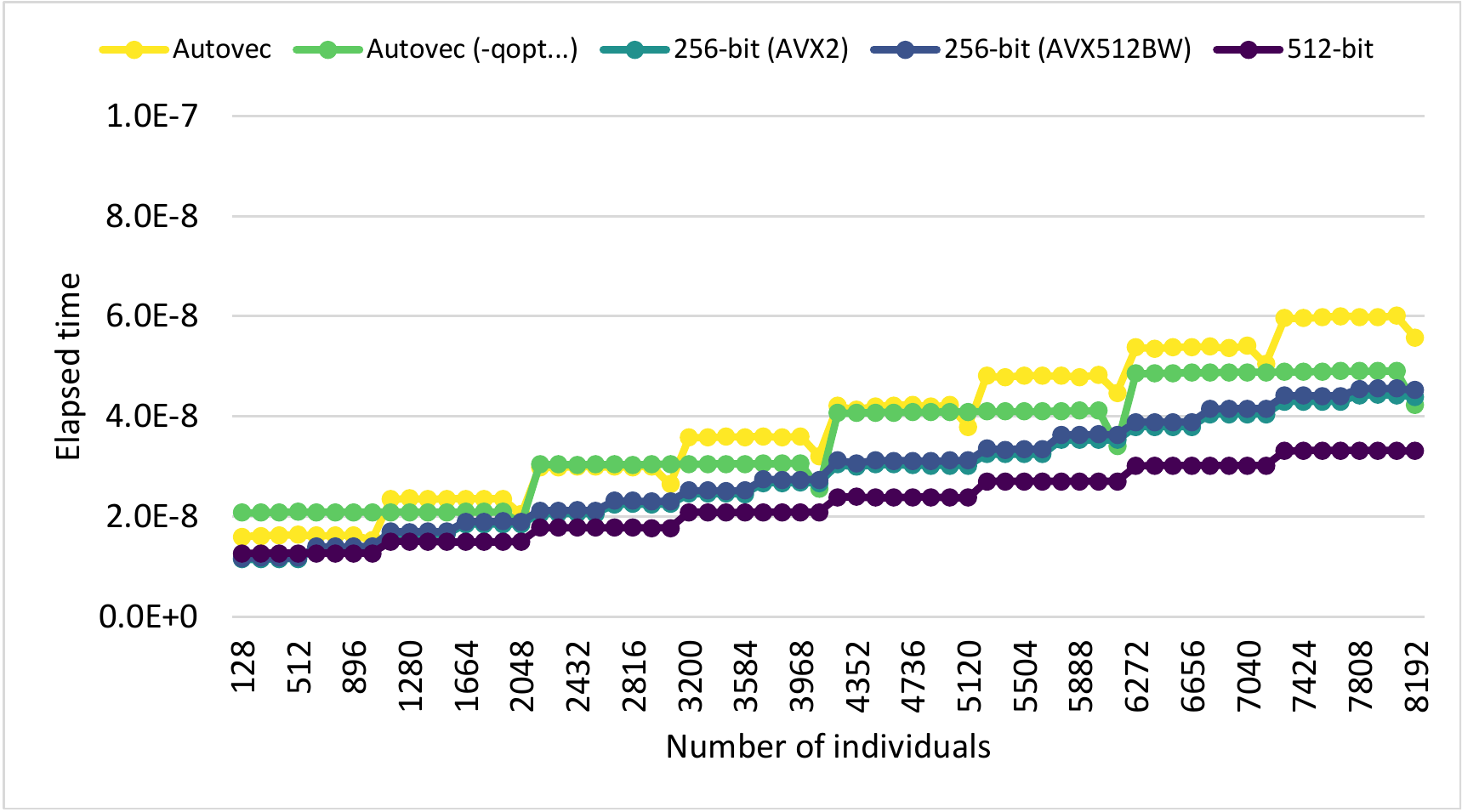}
  }
  \caption{
    Average elapsed time per cell during the exhaustive search of epistasis, for
    an increasing number of individuals, a fixed number of 680 SNPs, a fixed
    combination size of three and for both the GCC and Intel compilers.
  }\label{fig:search_individuals}
\end{figure*}

At last, Figs.~\ref{fig:search_order}~and~\ref{fig:search_individuals} present
the performance results for the whole exhaustive search algorithm. The two
figures compare the performance of the 256 and 512-bit explicit vectorization
approaches using operations from the \textit{AVX2} and \textit{AVX512BW}
extensions, with the automatically vectorized code using both GCC and Intel
compilers, for both versions of the search algorithm presented in
Algorithms~\ref{alg:nonseg}~and~\ref{alg:seg}.

Figs.~\ref{fig:search_order}~and~\ref{fig:search_individuals} use the average
elapsed time per cell to represent the performance results. The time per cell is
the average elapsed time spent during the computation of a single contingency
table cell, the subsequent MI operations corresponding to the cell of the table
and a fraction of the time spent during the calculations of previous genotype
tables (this time is equally divided across all cells of all combinations that
make use of that genotype table), for all of the contingency tables of the
order, number of SNPs and individuals specified.

Fig.~\ref{fig:search_order} represents the time per cell, shown as stacked bars
indicating the fraction of the time spent in each of the functions, during the
search of epistasis in combinations of 2, 4 and 8 SNPs, and for a fixed number
of individuals of 2048. The number of SNPs was tied to the size of the
combinations so that the workload among different explorations was as similar as
possible. Table~\ref{tbl:cells_per_order} indicates, for each exploration order,
the number of SNPs selected, the resulting number of SNP combinations of said
order, the number of different cells among those combinations and the difference
in workload that that exploration order and number of SNPs supposes from the
first one. The figure shows that the 512-bit explicit vectorization performs the
best out of all of the versions compared, which is coherent with what we saw
during the evaluation of the individual functions. The 256-bit explicit
implementations obtain practically the same results and the only
compiler-generated vectorization that beats any of the explicit vectorizations
is the Intel Compiler when coupled with the optimization flag
\texttt{-qopt-zmm-usage=high} for low-order epistasis searches.

The segmentation of operations introduced in the algorithm has an overall
positive effect on the explicitly vectorized implementations. From a CPU
frequency perspective, the segmentation algorithm achieves its goal. When there
is no separation between integer and binary arithmetic, and floating-point
arithmetic, the whole program runs at 2.8 and 2.5 GHz for the 256-bit and
512-bit implementations, respectively. However, when there is segmentation,
genotype and contingency table calculations run at 3.05 and 2.75 GHz, and the MI
operations run at 2.8 and 2.5 GHz for each implementation respectively. From the
performance perspective, the segmentation strategy only works with the
explicitly vectorized implementations and results in a noticeable reduction of
the elapsed time of the searches under GCC, and a much smaller gain under the
Intel Compiler for high-order interactions.

\begin{table}[t]
  \caption{
    Combination size and number of SNPs used during the exhaustive search
    evaluation, indicating the resulting number of combinations and table cells,
    and the relative workload deviation. The total number of cells is the
    product between the number of combinations (as indicated by the previous
    column) and the number of cells in each contingency table, \(3^{size}\), for
    a particular combination size. The deviation is the difference between the
    total number of cells for a particular combination size and SNP number, and
    the number of cells for combinations of two SNPs and 25000 SNPs, relative to
    the latter one.
  }
  \centering
  \small
  \begin{tabular}{lrrrrr}
    \toprule
    Size & SNPs  & Combinations & Total cells & Deviation \\
    \midrule
    2    & 25000 & 312487500    & 2812387500 & +0.00\%    \\
    4    & 242   & 34389810     & 2785574610 & -0.95\%    \\
    8    & 23    & 490314       & 3216950154 & +14.39\%   \\
    \bottomrule
  \end{tabular}\label{tbl:cells_per_order}
\end{table}

Fig.~\ref{fig:search_individuals} represents the time per cell for a growing
number of individuals from 128 to 8192, using a fixed combination size of three
and a fixed number of SNPs of 680. Although the number of individuals is
irrelevant to the calculation of the MI, it affects the calculation of the
genotype tables and contingency tables, and therefore the time per cell during
the whole search also grows linearly with the number of individuals, although
with a less pronounced slope than the two first individual operations. These two
subfigures show similar behaviour to the one shown in
Fig.~\ref{fig:ct_individuals} because the calculation of the contingency table
accounts for the majority of the elapsed time during the whole search.

Results from Fig.~\ref{fig:search_individuals} show that the explicit
vectorization using 512-bit operations achieve the best times. These are
very similar to those of the contingency table calculation function
(Fig.~\ref{fig:ct_individuals}), which makes sense since it is the most
time-consuming function of the algorithm as Fig.~\ref{fig:search_order} showed.
The best implementation is again the explicit 512-bit vectorization.

\subsection{Performance of the vectorized search compared against MPI3SNP}

To conclude the evaluation, Table~\ref{tbl:search_runtimes} compares the elapsed
time required to complete a third-order epistasis search using the original
MPI3SNP~\cite{ponte-fernandez_fast_2019} program and the explicit 512-bit
vectorization proposed in this paper, for an input data consisting of 1000 and
4000 SNPs and 1000 and 2000 individuals, and using a single core of the
processor. MPI3SNP was compiled using the same flags indicated during the
introduction of this section, enabling for both compilers the automatic
vectorization. Results show that the vector implementation of the algorithm
speeds the execution up by an average factor of 7 using GCC and 12 using the
Intel C++ Compiler. The speedup with the Intel Compiler is in part due to its
poor memory handling when allocating memory for objects inside a loop, something
that has been accounted for in the new implementation and that MPI3SNP does not
do. Therefore, we believe that the speedups obtained by GCC paint a more
realistic picture of what speedups should be expected of this algorithm.

\begin{table}
  \caption{
    Elapsed time, in seconds, and speedup of the explicitly vectorized 512-bit
    implementation compared to MPI3SNP\@. Results are the average of three
    executions of a single-threaded third-order search.
  }\label{tbl:search_runtimes}
  \centering
  \begin{tabular}{llllll}
    \toprule
      & \multirow{2}{*}{SNPs} & \multirow{2}{*}{Inds.}
      & MPI3SNP               & \multicolumn{2}{l}{512-bit vectorization} \\
      &                       &
      & Runtime               & Runtime & Speedup                         \\
    \midrule
    \multirow{4}{*}{\rotatebox[origin=c]{90}{GCC}}
      & 1000                  & 1000
      & 200.77                & 39.09   & 5.14                            \\
      & 1000                  & 2000
      & 254.43                & 50.97   & 4.99                            \\
      & 4000                  & 1000
      & 25698.86              & 2527.58 & 10.17                           \\
      & 4000                  & 2000
      & 31506.60              & 3401.53 & 9.26                            \\
    \midrule
    \multirow{4}{*}{\rotatebox[origin=c]{90}{Intel}}
      & 1000                  & 1000
      & 391.47                & 40.64   & 9.63                            \\
      & 1000                  & 2000
      & 735.52                & 48.98   & 15.02                           \\
      & 4000                  & 1000
      & 25113.33              & 2626.10 & 9.56                            \\
      & 4000                  & 2000
      & 47179.35              & 3288.37 & 14.35                           \\
    \bottomrule
  \end{tabular}
\end{table}

\section{Conclusions}\label{sec:conclusions}

Epistasis detection is still a biological problem that is far from being
resolved. Despite the efforts, there is no single method that can be applied to
a genome-wide scale data set to locate high order epistasis interactions
reliably~\cite{ponte-fernandez_evaluation_2020}. In this work, we propose
different SIMD implementations that exploit the parallelization opportunities
inherent to the epistasis detection problem in order to speed up the execution
of an exhaustive search. This is achieved by the introduction of AVX Intrinsics
functions during the calculation of the genotype tables, contingency tables and
MI metric. We also include general optimization strategies, such
as the segmentation of the operation pipeline due to the license-based
downclocking on Intel processors, and other optimizations specific to this code,
such as the loop unrolling during the calculation of genotype and contingency
tables, or the avoidance of logarithms of 0 during the MI calculation. Although
this work considers a specific exhaustive search algorithm, many of these
vectorization and optimization techniques could be directly applied to a
multitude of epistasis detection methods in the literature where genotype and
contingency tables are constructed to assess the association between a genotype
combination and a particular phenotype.

The results obtained highlight the potential of the SIMD parallelization when
applied to the epistasis detection problem. For example, the runtime under GCC
of an exhaustive search of an interaction consisting of three SNPs from two data
sets containing 4000 SNPs and 1000 and 2000 individuals, respectively, was
reduced from 428 and 525 minutes using MPI3SNP down to 42 and 57 minutes when
using the 512-bit vector implementation proposed in this work. The observed
speedups are not exclusive to single-core executions and will benefit
multi-threaded runs, accelerating the computations in each of the CPU cores
used.

The autovectorization provided by the compilers showed varying degrees of
success attending to the compiler and the operations considered. Intel, for
example, was capable of vectorizing all operations while GCC fell short. As for
the performance achieved, we observed that optimization flags play a big role in
the resulting performance of the code generated. GCC required the
\texttt{-fast-math} flag to be capable of vectorizing calls to the math library,
while Intel improved the performance significantly with the usage of the flag
\texttt{-qopt-zmm-usage=high}. With respect to performance, Intel's
autovectorization remained competitive with the explicit implementations for
low-order interaction searches but fell behind when moving past fourth-order
interactions. GCC's autovectorization, on the other hand, was never close to
the performance of the explicit implementations due to its failure of
vectorizing the operations.

Although the proposed explicit implementations are faster, they are tailored to
the x86\_64 CPU architecture. For different architectures, autovectorization is
the only alternative at the moment, and the results obtained in this study
suggest that compilers will be able to vectorize some computations of the
algorithm, if not all. If the vector width is comparable to those of AVX, as is
the case with ARM's Scalable Vector Extension, this algorithm is likely to be
effective. However, for architectures with much larger vector widths, e.g.\ the
NEC SX-Aurora TSUBASA vector processor, this algorithm will be inadequate, as
there are not enough values in the calculation of genotype and contingency
tables to fill the whole vector register.

Moving forward, with future CPU microarchitectures and the introduction of new
AVX extensions, it is reasonable to expect the performance of the SIMD epistasis
detection algorithm to improve even further. During the evaluation we saw the
effect that the Intel frequency model had on the performance attending to the
width of the operations, penalizing the larger vector widths. If these
differences in frequency are reduced in upcoming CPUs, the performance will
consequently increase. Furthermore, with future AVX instructions, for example,
the \textit{popcount} operation from the \textit{AVX512VPOPCNTDQ} extension
recently introduced with Intel Skylake processors, some operations of the
algorithm will allow for a more efficient implementation.

This work also presents some limitations and lines of future work:

\begin{enumerate}
  \item The algorithm described here is single-thread, and only focuses on the
        SIMD performance of a sequential execution. Combining the proposed SIMD
        implementations with a multi-thread or multi-process execution is
        fundamental for exploiting all the computing power that current CPUs and
        clusters of CPUs offer.

  \item This paper focuses solely on the x86\_64 architecture, the most extended
        architecture in general-purpose computers and high-performance clusters.
        Therefore, the proposed algorithm may or may not be appropriate for
        different CPU architectures, GPUs or vector processors such as the
        aforementioned NEC SX-Aurora TSUBASA\@. Studying different
        implementations of this algorithm, or a different algorithm better
        suited to the characteristics of another architecture, is something to
        consider.
\end{enumerate}

All codes presented in this work are available
online\footnote{https://github.com/UDC-GAC/fiuncho}.

\section*{Declaration of competing interest}

The authors declare that they have no known competing financial interests or
personal relationships that could have appeared to influence the work reported
in this paper.

\section*{Acknowledgments}

This work is supported by the Ministry of Science and Innovation of Spain
(PID2019{-}104184RB-I00 / AEI / 10.13039/\allowbreak{}501100011033), the Xunta
de Galicia and FEDER funds of the EU (Centro de Investigación de Galicia
accreditation 2019--2022, grant no. ED431G2019/01), Consolidation Program of
Competitive Research (grant no. ED431C 2017/04), and the FPU Program of the
Ministry of Education of Spain (grant no. FPU16/01333).

The authors would like to thank Agner Fog and Travis \allowbreak{} Downs, for
the invaluable information that they share in their personal webpages; Wojciech
Muła and all contributors of the \textit{sse-popcount} Github
repository\footnote{https://github.com/WojciechMula/sse-popcount}, for
maintaining the repository up to date which has saved us lots of time during
this work; and Marcos Horro, for listening to many rants during the development
and performance tuning of the AVX
vector codes.

\appendix
\renewcommand{\thelstlisting}{A.\arabic{lstlisting}}
\setcounter{lstlisting}{0}

\section{C++ functions}

\begin{lstlisting}[
  caption={Genotype table combination C++ function},
  label=lst:gt_base
]
inline void combine_subtable(
  const uint64_t *gt_tbl1, const size_t size1,
  const uint64_t *gt_tbl2, const size_t words,
  uint64_t *gt_tbl3)
{
  for (size_t i = 0; i < size1; i++) {*@\label{code:gt_3}@*
    for (size_t j = 0; j < 3; j++) {*@\label{code:gt_4}@*
      for (size_t k = 0; k < words; k++) {*@\label{code:gt_5}@*
        gt_tbl3[(i * 3 + j) * words + k] =*@\label{code:gt_6}@*
            gt_tbl1[i * words + k] &
            gt_tbl2[j * words + k];
      }
    }
  }
}

void combine(
  const GenotypeTable<uint64_t> &t1,*@\label{code:gt_1}@*
  const GenotypeTable<uint64_t> &t2,
  GenotypeTable<uint64_t> &out)*@\label{code:gt_2}@*
{
  combine_subtable(t1.cases, t1.size, t2.cases,
    t1.cases_words, out.cases);
  combine_subtable(t1.ctrls, t1.size, t2.ctrls,
    t1.ctrls_words, out.ctrls);
}
\end{lstlisting}

\begin{lstlisting}[
  caption={Contingency table calculation C++ function},
  label=lst:ct_base
]
inline void popcnt_subtable(
  const uint64_t *gt_tbl1, const size_t size1,
  const uint64_t *gt_tbl2, const size_t words,
  uint32_t *ct_tbl, const size_t ct_size)
{
  for (size_t i = 0; i < size1; i++) {
    for (size_t j = 0; j < 3; j++) {
      ct_tbl[i * 3 + j] = 0;
      for (size_t k = 0; k < words; k++) {
        ct_tbl[i * 3 + j] += std::bitset<64>(*@\label{code:ct_2}@*
          gt_tbl1[i * words + k] &
          gt_tbl2[j * words + k]).count();*@\label{code:ct_3}@*
      }
    }
  }
}

void combine_and_popcnt(
    const GenotypeTable<uint64_t> &t1,
    const GenotypeTable<uint64_t> &t2,
    ContingencyTable<uint32_t> &out)*@\label{code:ct_1}@*
{
  popcnt_subtable(t1.cases, t1.size, t2.cases,
    t1.cases_words, out.cases, out.size);
  popcnt_subtable(t1.ctrls, t1.size, t2.ctrls,
    t1.ctrls_words, out.ctrls, out.size);
}
\end{lstlisting}

\begin{lstlisting}[
  caption={MI computation C++ function},
  label=lst:mi_base
]
float MI(
  const ContingencyTable<uint32_t> &table,
  const float h_y,*@\label{code:mi_1}@*
  const float iinds)*@\label{code:mi_2}@*
{
  size_t i;
  float h_x = 0.0f, h_all = 0.0f;
  float p_case, p_ctrl, p_any;
  const size_t table_size = table.size;
  for (i = 0; i < table_size; i++) {
    p_case = table.cases[i] * iinds;*@\label{code:mi_3}@*
    if (p_case != 0.0f) {*@\label{code:mi_7}@*
      h_all -= p_case * logf(p_case);
    }
    p_ctrl = table.ctrls[i] * iinds;
    if (p_ctrl != 0.0f) {*@\label{code:mi_8}@*
      h_all -= p_ctrl * logf(p_ctrl);
    }*@\label{code:mi_4}@*
    p_any = p_case + p_ctrl;*@\label{code:mi_5}@*
    if (p_any != 0.0f) {*@\label{code:mi_9}@*
      h_x -= p_any * logf(p_any);
    }*@\label{code:mi_6}@*
  }
  return h_x + h_y - h_all;
}
\end{lstlisting}

\begin{lstlisting}[
  caption={Genotype table calculation auxiliary C++ function vectorized with
    \textit{AVX2} Intrinsics},
  label=lst:gt_avx2
]
inline void combine_subtable(
  const uint64_t *gt_tbl1, const size_t size1,
  const uint64_t *gt_tbl2, const size_t words,
  uint64_t *gt_tbl3)
{
  size_t i, j, k;
  const __m256i *ptr1 = gt_tbl1;
  for (i = 0; i < size1; ++i) {
    const __m256i *ptr2_1 = gt_tbl2 + 0 * words;
    const __m256i *ptr2_2 = gt_tbl2 + 1 * words;
    const __m256i *ptr2_3 = gt_tbl2 + 2 * words;
    __m256i *ptr3_1 = gt_tbl3 + (i*3+0) * words;
    __m256i *ptr3_2 = gt_tbl3 + (i*3+1) * words;
    __m256i *ptr3_3 = gt_tbl3 + (i*3+2) * words;
    for (k = 0; k < words; k += 4) {
      __m256i y0 = _mm256_load_si256(ptr1++);*@\label{code:vgt_1}@*
      __m256i y1 = _mm256_load_si256(ptr2_1++);
      __m256i y2 = _mm256_load_si256(ptr2_2++);
      __m256i y3 = _mm256_load_si256(ptr2_3++);
      _mm256_store_si256(ptr3_1++,*@\label{code:vgt_2}@*
        _mm256_and_si256(y0, y1));
      _mm256_store_si256(ptr3_2++,
        _mm256_and_si256(y0, y2));
      _mm256_store_si256(ptr3_3++,
        _mm256_and_si256(y0, y3));
    }
  }
}
\end{lstlisting}

\begin{lstlisting}[
  caption={Contingency table calculation auxiliary C++ function vectorized with
    \textit{AVX2} Intrinsics},
  label={lst:ct_avx2}
]
inline void iter(
  const uint64_t *ptr1, const uint64_t *ptr2,
  const __m256i &lu, const __m256i &low_mask,
  __m256i &local)
{
  __m256i o1 = _mm256_load_si256(ptr1);*@\label{code:vct256_5}@*
  __m256i o2 = _mm256_load_si256(ptr2);*@\label{code:vct256_6}@*
  __m256i vec = _mm256_and_si256(o1, o2);*@\label{code:vct256_7}@*
  __m256i lo = _mm256_and_si256(vec, low_mask);*@\label{code:vct256_8}@*
  __m256i hi = _mm256_and_si256(
    _mm256_srli_epi16(vec, 4), low_mask);
  __m256i popcnt1 = _mm256_shuffle_epi8(lu, lo);
  __m256i popcnt2 = _mm256_shuffle_epi8(lu, hi);
  local = _mm256_add_epi8(local, popcnt1);
  local = _mm256_add_epi8(local, popcnt2);*@\label{code:vct256_9}@*
}

inline void popcnt_subtable(
  const uint64_t *gt_tbl1, const size_t size1,
  const uint64_t *gt_tbl2, const size_t words,
  uint32_t *ct_tbl, const size_t ct_size)
{
  const __m256i lookup = _mm256_setr_epi8(
    /* 0 */ 0, /* 1 */ 1, /* 2 */ 1, /* 3 */ 2,
    /* 4 */ 1, /* 5 */ 2, /* 6 */ 2, /* 7 */ 3,
    /* 8 */ 1, /* 9 */ 2, /* a */ 2, /* b */ 3,
    /* c */ 2, /* d */ 3, /* e */ 3, /* f */ 4,
    /* 0 */ 0, /* 1 */ 1, /* 2 */ 1, /* 3 */ 2,
    /* 4 */ 1, /* 5 */ 2, /* 6 */ 2, /* 7 */ 3,
    /* 8 */ 1, /* 9 */ 2, /* a */ 2, /* b */ 3,
    /* c */ 2, /* d */ 3, /* e */ 3, /* f */ 4);
  const __m256i low_mask = _mm256_set1_epi8(0xf);

  size_t i, j, k;
  for (i = 0; i < size1; ++i) {*@\label{code:vct256_1}@*
    for (j = 0; j < 3; ++j) {*@\label{code:vct256_2}@*
      __m256i acc = _mm256_setzero_si256();
      for (k = 0; k + 32 <= words; k += 32) {*@\label{code:vct256_3}@*
        __m256i local = _mm256_setzero_si256();
        iter(gt_tbl1 + i * words + k + 0,
          gt_tbl2 + j * words + k + 0,
          lookup, low_mask, local);
        iter(gt_tbl1 + i * words + k + 4,
          gt_tbl2 + j * words + k + 4,
          lookup, low_mask, local);
        iter(gt_tbl1 + i * words + k + 8,
          gt_tbl2 + j * words + k + 8,
          lookup, low_mask, local);
        iter(gt_tbl1 + i * words + k + 12,
          gt_tbl2 + j * words + k + 12,
          lookup, low_mask, local);
        iter(gt_tbl1 + i * words + k + 16,
          gt_tbl2 + j * words + k + 16,
          lookup, low_mask, local);
        iter(gt_tbl1 + i * words + k + 20,
          gt_tbl2 + j * words + k + 20,
          lookup, low_mask, local);
        iter(gt_tbl1 + i * words + k + 24,
          gt_tbl2 + j * words + k + 24,
          lookup, low_mask, local);
        iter(gt_tbl1 + i * words + k + 28,
          gt_tbl2 + j * words + k + 28,
          lookup, low_mask, local);
        acc = _mm256_add_epi64(acc,
          _mm256_sad_epu8(local,
            _mm256_setzero_si256()));
      }*@\label{code:vct256_10}@*

      __m256i local = _mm256_setzero_si256();
      for (; k < words; k += 4) {
        iter(gt_tbl1 + i * words + k,
          gt_tbl2 + j * words + k,
          lookup, low_mask, local);
      }*@\label{code:vct256_4}@*
      acc = _mm256_add_epi64(acc,
        _mm256_sad_epu8(local,
          _mm256_setzero_si256()));

      ct_tbl[i * 3 + j] =
        _mm256_extract_epi64(acc, 0) +
        _mm256_extract_epi64(acc, 1) +
        _mm256_extract_epi64(acc, 2) +
        _mm256_extract_epi64(acc, 3);
    }
  }
  for (i = size1 * 3; i < ct_size; ++i) {
      ct_tbl[i] = 0;
  }
}
\end{lstlisting}

\begin{lstlisting}[
  caption={Contingency table calculation auxiliary C++ function vectorized with
    \textit{AVX512BW} Intrinsics
  },
  label=lst:ct_avx512bw
]
inline void popcnt_subtable(
  const uint64_t *gt_tbl1, const size_t size1,
  const uint64_t *gt_tbl2, const size_t words,
  uint32_t *ct_tbl, const size_t ct_size)
{
  const __m512i lookup = _mm512_setr_epi64(
    0x0302020102010100llu,0x0403030203020201llu,
    0x0302020102010100llu,0x0403030203020201llu,
    0x0302020102010100llu,0x0403030203020201llu,
    0x0302020102010100llu,0x0403030203020201llu);
  const __m512i low_mask = _mm512_set1_epi8(0xf);

  size_t i, j, k, l;
  for (i = 0; i < size1; ++i) {*@\label{code:vct512_1}@*
    for (j = 0; j < 3; ++j) {*@\label{code:vct512_2}@*
      k = 0;
      __m512i acc = _mm512_setzero_si512();
      while (k < words) {*@\label{code:vct512_3}@*
        __m512i local = _mm512_setzero_si512();
        for (l = 0; l < 255 / 8 && k < words;
              ++l, k += 8) {
          __m512i z0 = _mm512_load_si512(*@\label{code:vct512_4}@*
            gt_tbl2 + j * words + k);
          __m512i z1 = _mm512_load_si512(*@\label{code:vct512_5}@*
            gt_tbl1 + i * words + k);
          __m512i z2 = _mm512_and_si512(z0, z1);*@\label{code:vct512_6}@*
          __m512i lo = _mm512_and_si512(
            z2, low_mask);
          __m512i hi = _mm512_and_si512(
            _mm512_srli_epi32(z2, 4), low_mask);
          __m512i popcnt1 = _mm512_shuffle_epi8(
            lookup, lo);
          __m512i popcnt2 = _mm512_shuffle_epi8(
            lookup, hi);
          local = _mm512_add_epi8(local,popcnt1);
          local = _mm512_add_epi8(local,popcnt2);
        }
        acc = _mm512_add_epi64(acc,
          _mm512_sad_epu8(local,
            _mm512_setzero_si512()));
      }
      ct_tbl[i * 3 + j] =
        _mm512_reduce_add_epi64(acc);
    }
  }
  for (i = size1 * 3; i < ct_size; ++i) {
    ct_tbl[i] = 0;
  }
}
\end{lstlisting}

\begin{lstlisting}[
  caption={
    MI computation C++ function vectorized with \textit{AVX2} Intrinsics
  },
  label=lst:mi_avx2
]
float MI(
  const ContingencyTable<uint32_t> &table,
  const float h_y,
  const float iinds)
{
  const __m256 ones = _mm256_set1_ps(1.0);
  const __m256 ii = _mm256_set1_ps(iinds);
  __m256 h_x = _mm256_setzero_ps();
  __m256 h_all = _mm256_setzero_ps();
  __m256i y0;
  __m256 y1, y2, y3, y4, y5;
  for (auto i = 0; i < table.size; i += 8) {
    y0 = _mm256_load_si256(table.cases + i);
    y3 = _mm256_mul_ps(_mm256_cvtepi32_ps(y0), ii);
    // Identify cells with 0's
    y1 = _mm256_cmp_ps(y0,*@\label{code:vmi256_1}@*
      _mm256_setzero_ps(), _CMP_NEQ_OQ);
    // Replace 0's with 1's before log
    y4 = _mm256_log_ps(_mm256_blendv_ps(
      ones, y3, y1));*@\label{code:vmi256_2}@*
    h_all = _mm256_fmadd_ps(y3, y4, h_all);
    y0 = _mm256_load_si256(table.ctrls + i);
    y4 = _mm256_mul_ps(_mm256_cvtepi32_ps(y0), ii);
    // Identify cells with 0's
    y2 = _mm256_cmp_ps(y0,*@\label{code:vmi256_3}@*
      _mm256_setzero_ps(), _CMP_NEQ_OQ);
    // Replace 0's with 1's before log
    y5 = _mm256_log_ps(_mm256_blendv_ps(
      ones, y4, y2));*@\label{code:vmi256_4}@*
    h_all = _mm256_fmadd_ps(y4, y5, h_all);
    y5 = _mm256_add_ps(y3, y4);
    // Merge previous masks
    y1 = _mm256_or_ps(y1, y2);*@\label{code:vmi256_5}@*
    // Replace 0's with 1's before log
    y3 = _mm256_log_ps(_mm256_blendv_ps(
      ones, y5, y1));*@\label{code:vmi256_6}@*
    h_x = _mm256_fmadd_ps(y5, y3, h_x);
  }
  y3 = _mm256_hadd_ps(h_all, h_x);
  return (y3[0] + y3[1] + y3[4] + y3[5]) -
    h_y - (y3[2] + y3[3] + y3[6] + y3[7]);
}
\end{lstlisting}

\begin{lstlisting}[
  caption={
    MI computation C++ function vectorized with \textit{AVX512BW} Intrinsics
  },
  label=lst:mi_avx512f512
]
float MI(
  const ContingencyTable<uint32_t> &table,
  const float h_y,
  const float iinds)
{
  const __m512 ones = _mm512_set1_ps(1.0);
  const __m512 ii = _mm512_set1_ps(inv_inds);
  __m512 h_x = _mm512_setzero_ps();
  __m512 h_all = _mm512_setzero_ps();
  __m512i z0;
  __m512 z1, z2, z3, z4;
  __mmask16 mask1, mask2, mask3;
  for (auto i = 0; i < table.size; i += 8) {
    z0 = _mm512_load_si512(table.cases + i);
    z2 = _mm512_mul_ps(
      _mm512_cvtepi32_ps(z0), ii);
    // Identify cells with 0's
    mask1 = _mm512_cmp_epi32_mask(z0,*@\label{code:vmi512_1}@*
      _mm512_setzero_si512(), _MM_CMPINT_NE);
    // Replace 0's with 1's before log
    z3 = _mm512_log_ps(_mm512_mask_blend_ps(*@\label{code:vmi512_4}@*
      mask1, ones, z2));
    h_all = _mm512_fmadd_ps(z2, z3, h_all);
    z0 = _mm512_load_si512(table.ctrls + i);
    z3 = _mm512_mul_ps(
      _mm512_cvtepi32_ps(z0), ii);
    // Identify cells with 0's
    mask2 = _mm512_cmp_epi32_mask(z0,*@\label{code:vmi512_2}@*
      _mm512_setzero_si512(), _MM_CMPINT_NE);
    // Replace 0's with 1's before log
    z4 = _mm512_log_ps(_mm512_mask_blend_ps(*@\label{code:vmi512_5}@*
      mask2, ones, z3));
    h_all = _mm512_fmadd_ps(z3, z4, h_all);
    z1 = _mm512_add_ps(z2, z3);
    // Merge previous masks
    mask3 = _kor_mask16(mask1, mask2);*@\label{code:vmi512_3}@*
    // Replace 0's with 1's before log
    z2 = _mm512_log_ps(_mm512_mask_blend_ps(*@\label{code:vmi512_6}@*
      mask3, ones, z1));
    h_x = _mm512_fmadd_ps(z1, z2, h_x);
  }

  return _mm512_reduce_add_ps(h_all) -
    _mm512_reduce_add_ps(h_x) - h_y;
}
\end{lstlisting}

\bibliographystyle{elsarticle-num}
\bibliography{main}

\begin{thebibliography}{10}
\expandafter\ifx\csname url\endcsname\relax
  \def\url#1{\texttt{#1}}\fi
\expandafter\ifx\csname urlprefix\endcsname\relax\def\urlprefix{URL }\fi
\expandafter\ifx\csname href\endcsname\relax
  \def\href#1#2{#2} \def\path#1{#1}\fi

\bibitem{churchill_epistasis_2013}
G.~Churchill, Epistasis, in: S.~Maloy, K.~Hughes (Eds.), Brenner's
  {Encyclopedia} of {Genetics} ({Second} {Edition}), second edition Edition,
  Academic Press, San Diego, 2013, pp. 505--507.
\newblock \href
  {https://doi.org/https://doi.org/10.1016/B978-0-12-374984-0.00482-4}
  {\path{doi:https://doi.org/10.1016/B978-0-12-374984-0.00482-4}}.

\bibitem{he_genome_2017}
S.~He, J.~C. Reif, V.~Korzun, R.~Bothe, E.~Ebmeyer, Y.~Jiang, Genome-wide
  mapping and prediction suggests presence of local epistasis in a vast elite
  winter wheat populations adapted to {Central} {Europe}, Theoretical and
  Applied Genetics 130~(4) (2017) 635--647.

\bibitem{jiang_quantitative_2017}
Y.~Jiang, R.~H. Schmidt, Y.~Zhao, J.~C. Reif, A quantitative genetic framework
  highlights the role of epistatic effects for grain-yield heterosis in bread
  wheat, Nature genetics 49~(12) (2017) 1741--1746.

\bibitem{banerjee_genome_2020}
P.~Banerjee, V.~A.~O. Carmelo, H.~N. Kadarmideen, Genome-{Wide} {Epistatic}
  {Interaction} {Networks} {Affecting} {Feed} {Efficiency} in {Duroc} and
  {Landrace} {Pigs}, Frontiers in genetics 11 (2020) 121.

\bibitem{ruiz_evidence_2017}
O.~Ruiz-Larra{\~n}aga, P.~V{\'a}zquez, M.~Iriondo, C.~Manzano, M.~Aguirre,
  J.~M. Garrido, R.~A. Juste, A.~Estonba, Evidence for gene-gene epistatic
  interactions between susceptibility genes for {Mycobacterium} avium subsp.
  paratuberculosis infection in cattle, Livestock Science 195 (2017) 63--66.

\bibitem{meijsen_using_2018}
J.~J. Meijsen, A.~Rammos, A.~Campbell, C.~Hayward, D.~J. Porteous, I.~J. Deary,
  R.~E. Marioni, K.~K. Nicodemus, Using tree-based methods for detection of
  gene–gene interactions in the presence of a polygenic signal: simulation
  study with application to educational attainment in the {Generation}
  {Scotland} {Cohort} {Study}, Bioinformatics 35~(2) (2018) 181--188.
\newblock \href {https://doi.org/https://doi.org/10.1093/bioinformatics/bty462}
  {\path{doi:https://doi.org/10.1093/bioinformatics/bty462}}.

\bibitem{wollstein_novel_2017}
A.~Wollstein, S.~Walsh, F.~Liu, U.~Chakravarthy, M.~Rahu, J.~H. Seland,
  G.~Soubrane, L.~Tomazzoli, F.~Topouzis, J.~R. Vingerling, et~al., Novel
  quantitative pigmentation phenotyping enhances genetic association,
  epistasis, and prediction of human eye colour, Scientific reports 7~(1)
  (2017) 1--11.

\bibitem{kim_towards_2020}
Y.-H. Kim, Y.~Yoon, Y.-H. Kim, Towards a {Better} {Basis} {Search} through a
  {Surrogate} {Model}-{Based} {Epistasis} {Minimization} for {Pseudo}-{Boolean}
  {Optimization}, Mathematics 8~(8) (2020) 1287.

\bibitem{shang_epiminer_2014}
J.~Shang, J.~Zhang, Y.~Sun, Y.~Zhang, {EpiMiner}: {A} three-stage
  co-information based method for detecting and visualizing epistatic
  interactions, Digital Signal Processing 24 (2014) 1--13.

\bibitem{sun_epiaco_2017}
Y.~Sun, J.~Shang, J.-X. Liu, S.~Li, C.-H. Zheng, {epiACO} - a method for
  identifying epistasis based on ant {Colony} optimization algorithm, BioData
  mining 10~(1) (2017) 1--17.

\bibitem{wang_bayesian_2015}
J.~Wang, T.~Joshi, B.~Valliyodan, H.~Shi, Y.~Liang, H.~T. Nguyen, J.~Zhang,
  D.~Xu, A {Bayesian} model for detection of high-order interactions among
  genetic variants in genome-wide association studies, Bmc Genomics 16~(1)
  (2015) 1--20.

\bibitem{ponte-fernandez_evaluation_2020}
C.~Ponte-Fernandez, J.~Gonzalez-Dominguez, A.~Carvajal-Rodriguez, M.~J. Martin,
  Evaluation of {Existing} {Methods} for {High}-{Order} {Epistasis}
  {Detection}, IEEE/ACM Transactions on Computational Biology and
  Bioinformatics (2020).
\newblock \href {https://doi.org/https://doi.org/10.1109/TCBB.2020.3030312}
  {\path{doi:https://doi.org/10.1109/TCBB.2020.3030312}}.

\bibitem{wan_boost_2010}
X.~Wan, C.~Yang, Q.~Yang, H.~Xue, X.~Fan, N.~L. Tang, W.~Yu, {BOOST}: {A}
  {Fast} {Approach} to {Detecting} {Gene}-{Gene} {Interactions} in
  {Genome}-wide {Case}-{Control} {Studies}, The American Journal of Human
  Genetics 87~(3) (2010) 325--340.
\newblock \href {https://doi.org/https://doi.org/10.1016/j.ajhg.2010.07.021}
  {\path{doi:https://doi.org/10.1016/j.ajhg.2010.07.021}}.

\bibitem{campos_heterogeneous_2020}
R.~Campos, D.~Marques, S.~Santander-Jim{\'e}nez, L.~Sousa, A.~Ilic,
  Heterogeneous {CPU}+{iGPU} {Processing} for {Efficient} {Epistasis}
  {Detection}, in: M.~Malawski, K.~Rzadca (Eds.), Euro-Par 2020: Parallel
  Processing, Springer International Publishing, 2020, pp. 613--628.

\bibitem{martinez_fast_2018}
H.~Mart{\'\i}nez, S.~Barrachina, M.~Castillo, E.~S. Quintana-Orti, J.~Rambla~de
  Argila, X.~Farr{\'e}, A.~Navarro, {FaST}-{LMM} for {Two}-{Way} {Epistasis}
  {Tests} on {High}-{Performance} {Clusters}, Journal of Computational Biology
  25~(8) (2018) 862--870.

\bibitem{ponte-fernandez_fast_2019}
C.~Ponte-Fernández, J.~González-Domínguez, M.~J. Martín, Fast search of
  third-order epistatic interactions on {CPU} and {GPU} clusters, The
  International Journal of High Performance Computing Applications (2019)
  20--29\href {https://doi.org/https://doi.org/10.1177/1094342019852128}
  {\path{doi:https://doi.org/10.1177/1094342019852128}}.

\bibitem{gonzalez_gpu_2015}
J.~Gonz{\'a}lez-Dom{\'\i}nguez, B.~Schmidt, Gpu-accelerated exhaustive search
  for third-order epistatic interactions in case--control studies, Journal of
  Computational Science 8 (2015) 93--100.

\bibitem{nobre_exploring_2020}
R.~{Nobre}, A.~{Ilic}, S.~{Santander-Jiménez}, L.~{Sousa}, Exploring the
  {Binary} {Precision} {Capabilities} of {Tensor} {Cores} for {Epistasis}
  {Detection}, in: 2020 IEEE International Parallel and Distributed Processing
  Symposium (IPDPS), 2020, pp. 338--347.
\newblock \href {https://doi.org/https://doi.org/10.1109/IPDPS47924.2020.00043}
  {\path{doi:https://doi.org/10.1109/IPDPS47924.2020.00043}}.

\bibitem{wienbrandt_fpga_2014}
L.~Wienbrandt, J.~C. K{\"a}ssens, J.~Gonz{\'a}lez-Dom{\'\i}nguez, B.~Schmidt,
  D.~Ellinghaus, M.~Schimmler, {FPGA}-based {Acceleration} of {Detecting}
  {Statistical} {Epistasis} in {GWAS}, Procedia Computer Science 29 (2014)
  220--230.

\bibitem{luecke_fast_2015}
G.~R. {Luecke}, N.~T. {Weeks}, B.~M. {Groth}, M.~{Kraeva}, L.~{Ma}, L.~M.
  {Kramer}, J.~E. {Koltes}, J.~M. {Reecy}, Fast {Epistasis} {Detection} in
  {Large-Scale} {GWAS} for {Intel} {Xeon} {Phi} {Clusters}, in: 2015 IEEE
  Trustcom/BigDataSE/ISPA, Vol.~3, 2015, pp. 228--235.
\newblock \href {https://doi.org/https://doi.org/10.1109/Trustcom.2015.637}
  {\path{doi:https://doi.org/10.1109/Trustcom.2015.637}}.

\bibitem{galvez_blvector_2021}
S.~G{\'a}lvez, F.~Agostini, J.~Caselli, P.~Hernandez, G.~Dorado, {BLVector}:
  {Fast} {BLAST}-like algorithm for manycore {CPU} with vectorization,
  Frontiers in Genetics 12 (2021).

\bibitem{rucci_swimm_2019}
E.~Rucci, C.~G. Sanchez, G.~B. Juan, A.~De~Giusti, M.~Naiouf, M.~Prieto-Matias,
  {SWIMM} 2.0: enhanced {Smith}--{Waterman} on {Intel}’s multicore and
  manycore architectures based on {AVX}-512 vector extensions, International
  Journal of Parallel Programming 47~(2) (2019) 296--316.

\bibitem{yin_rabbitmash_2020}
Z.~Yin, X.~Xu, J.~Zhang, Y.~Wei, B.~Schmidt, W.~Liu, {RabbitMash}:
  {Accelerating} hash-based genome analysis on modern multi-core architectures,
  Bioinformatics (2020).

\bibitem{guo_cloud_2014}
X.~Guo, Y.~Meng, N.~Yu, Y.~Pan, Cloud computing for detecting high-order
  genome-wide epistatic interaction via dynamic clustering, BMC bioinformatics
  15~(1) (2014) 1--16.

\bibitem{mula_faster_2018}
W.~Mu{\l}a, N.~Kurz, D.~Lemire, Faster population counts using avx2
  instructions, The Computer Journal 61~(1) (2018) 111--120.

\bibitem{mula_github_sse}
W.~Muła, {GitHub} repository "{SIMD} popcount", containing the vector
  algorithms published in \cite{mula_faster_2018},
  \url{https://github.com/WojciechMula/sse-popcount}, accessed: 2021-02-24
  (2016).

\bibitem{intel_xeon_specification}
I.~Corporation, Second {Generation} {Intel} {Xeon} {Scalable} {Processors}
  {Specification} {Update},
  \url{https://www.intel.com/content/dam/www/public/us/en/documents/specification-updates/2nd-gen-xeon-scalable-spec-update.pdf},
  accessed: 2020-11-07 (September 2020).

\end{thebibliography}

\end{document}